\documentclass[%
  reprint,
  superscriptaddress,
  nofootinbib,
  amsmath,amssymb,
  aps,
  floatfix,
  prd
]{revtex4-1}

\usepackage{graphicx}
\usepackage{bm}

\usepackage{microtype,booktabs}
\usepackage[dvipsnames]{xcolor}
\usepackage[utf8]{inputenc}
\def\fm {\mathop{\hbox{fm}}}
\def\MeV {\mathop{\hbox{MeV}}}

\def\beq{\begin{equation}}
\def\eeq{\end{equation}}
\def\beqs#1\eeqs{\beq\begin{split} #1 \end{split}\eeq}

\long\def\comment#1{}

\def\av#1{ \left\langle #1 \right\rangle }

\usepackage[colorlinks=true,backref=false,linktocpage=true,
  citecolor=NavyBlue,urlcolor=NavyBlue,linkcolor=NavyBlue,pdfpagemode=UseOutlines]{hyperref}
\hypersetup{%
  bookmarksnumbered=true,
  pdftitle = {},
  pdfsubject = {},
  pdfauthor = {},
  pdfkeywords = {}
}

\begin{document}

\title{Charged pion electric polarizability from lattice QCD}

\author{Hossein Niyazi}
\email{hosseinniyazi@gwu.edu}
\affiliation{Department of Physics, The George Washington University, Washington, DC 20052, USA}
\author{Andrei Alexandru}%
\email{aalexan@gwu.edu}
\affiliation{Department of Physics, The George Washington University, Washington, DC 20052, USA}
\affiliation{Department of Physics, University of Maryland, College Park, MD 20742, USA}

\author{Frank X. Lee}
\email{fxlee@gwu.edu}
\affiliation{Department of Physics, The George Washington University, Washington, DC 20052, USA}

\author{Michael Lujan}
\email{mlujan@lmi.org}
\affiliation{Advanced Analytics \& Artificial Intelligence Group, LM\~I, Tysons Corner, VA 22102, USA}


\begin{abstract}
We present a calculation of the charged pion electric polarizability using the background field method. To extract the mass-shift induced by the electric field for the accelerated charged particle we fit the lattice QCD correlators using correlators derived from an effective model. The methodology outlined in this study~(boundary conditions, fitting procedure, etc) is designed to ensure that the results are invariant under gauge transformations of the background field. 
We apply the method to four $N_f=2$ dynamical ensembles to extract  $\alpha_{\pi^\pm}$ at pion mass of $315 \MeV$.
\end{abstract}

\maketitle


\section{\label{sec:intro}Introduction}

Electromagnetic polarizabilities are important properties that shed light on the internal structure of hadrons. The charged quarks inside a hadron respond to applied electromagnetic fields, revealing the deformation (or `stiffness') of the composite system as a fraction of its volume for uniform fields, and charge and current distributions for varying fields. 
There is an active community in nuclear physics partaking in this endeavor.
Experimentally, polarizabilities are primarily studied by low-energy Compton scattering. On the theoretical side, a variety of methods have been employed to describe the physics involved, from dispersion relations~\cite{Lvov:1993fp}, to chiral perturbation theory (ChPT)~\cite{Moinester:2019sew,Lensky:2009uv,Hagelstein:2020vog} or chiral effective field theory (EFT)~\cite{Griesshammer:2012we},
 to lattice QCD. Refs.~\cite{Moinester:2019sew,Griesshammer:2012we} also contain reviews of the experimental status.

Understanding electromagnetic polarizabilities has been a long-term goal of lattice QCD. One challenge is that it requires the application of both QCD and QED first principles. The standard tool to compute polarizabilities is the background field method~\cite{Fiebig:1988en,Lujan:2016ffj, Lujan:2014kia, Freeman:2014kka, Freeman:2013eta, Tiburzi:2008ma, Detmold:2009fr, Alexandru:2008sj, Lee:2005dq, Lee:2005ds,Engelhardt:2007ub,Bignell:2020xkf,Deshmukh:2017ciw,Bali:2017ian,Bruckmann:2017pft,Parreno:2016fwu,Luschevskaya:2015cko,Chang:2015qxa,Detmold:2010ts}. Although such calculations are relatively straightforward, the challenge lies in determining very small energy shift relative to the mass of the hadron. An alternative is based on evaluating derivatives with respect to the external field directly, which requires evaluation of
 four-point functions~(current-current correlators). This directly mimics the Compton scattering process on the lattice, but is significantly more demanding computationally~\cite{Andersen:1996qb,Wilcox:1996vx}. Methods to study higher-order polarizabilities have also been proposed~\cite{Davoudi:2015cba,Engelhardt:2011qq,Lee:2011gz,Detmold:2006vu}.

In this work, we focus on the electric polarizability of charged pions in the background field method, which presents its own challenge because the entire system accelerates in the presence of electric field. This motion is unrelated to polarizability and must be isolated from the deformation due to quark and gluon dynamics inside the hadron. Standard plateau technique of extracting energy from the large-time behavior of the two-point correlator fails. Our goal is to find a robust methodology to take the motion into account so that the polarizability can be reliably extracted. In this study we employ Dirichlet boundary conditions and address a number of related theoretical issues.

In Section~\ref{sec:effectivemodel}, we discuss the effective model we use to capture the behavior of the charged pion. 
In Section~\ref{sec:fittingprodecure}, we describe our fitting procedure incorporating the effective model and present results on the polarizability from lattice QCD simulations. Conclusions are given in Section~\ref{sec:con}.

\section{\label{sec:effectivemodel} Methodology}

\subsection{Background field method}
To extract the polarizability, we compute the pion mass shift induced by an external electric field. For technical convenience we use imaginary electric fields and the mass shift is then
\beq
\Delta m= \frac12 \alpha \mathcal{E}^2.
\label{eq:shift}
\eeq
The {\em imaginary} background electric field is introduced by multiplying the $SU(3)$ gauge links by the $U(1)$ links
\beq
    U_{\mu}(x) = e^{-iq a A_\mu(x)}.
    \label{eq:U1}
\eeq
Note that here the charge $q$ is the charge of the relevant quark field: for down quarks $q_d=-e/3$ and for up quarks $q_u=2e/3$,
with $e$ the absolute value for the electron charge.
There are an infinite number of $U(1)$ gauges that produce the same electric field; one such gauge is the $t$-dependent vector potential on the $x$-links,
\beqs
	A_\mu(x,t) &=  {\cal E} (t-t_0) \, \delta_{\mu,x},
	\label{eq:gauge}
\eeqs
where $t_0$ is the origin of the potential. Since the calculation will be performed in a gauge-invariant manner, the value of $t_0$ can be shifted via a gauge transformation; we set $t_0=0$. The potential produces a constant electric field in the $x$ direction except at the edges of the lattice. 

%
%

Note that the interaction with the background field is physical only
if the electric field is real. In this case the electric field leads
to a real-valued exponential factor in Eq.~\eqref{eq:U1}.
Using an imaginary electric field is convenient since in this case
the factor in Eq.~\eqref{eq:U1} is a complex factor of modulus one 
and the implementation for the discretizied Dirac operator is
basically unchanged. Using imaginary fields is justified because 
the theory, with Dirichlet boundary conditions, is analytic for 
small values of ${\cal E}$ and we can use analytically continuation 
to make the fields imaginary.
The two formulations are equivalent as long as analytic continuation is valid and the sign is corrected in the interpretation of Eq.~\eqref{eq:shift} (see discussion of the issue in Refs.~\cite{Alexandru:2008sj,Shintani:2006xr}).
We note in passing that there is no such ambiguity for introducing a constant magnetic field on the lattice where only spatial coordinates are involved. A real magnetic field leads to a $U(1)$ phase factor and the magnetic polarizability term in the mass shift has the negative sign: $\Delta m=-\beta B^2/2$.

\subsection{Relativistic effective correlator}
In the presence of the background field the two-point correlator for charged particles is not expected to be a simple sum of exponentials, as is the case for the neutral particles. A work around is to fit the hadron correlators against the functional form expected for a charged particle propagator in the presence of an electric field~\cite{Detmold:2009dx}. As it turns out the proposed correlation function computed in the infinite-volume is not a good fit for our correlators since the finite volume effects are important. We proposed a method that incorporates these effects~\cite{Alexandru:2015dva}. In this section we outline the method used to compute the fit correlator.
We show that our correlator converges in the infinite volume limit to the expected correlator and that the finite volume effects are important.

\begin{figure*}[t]
    \centering
    \begin{tabular}{cc}
    \parbox[c]{8.5cm}{\includegraphics[height=6cm,trim=0cm 0 0 0]{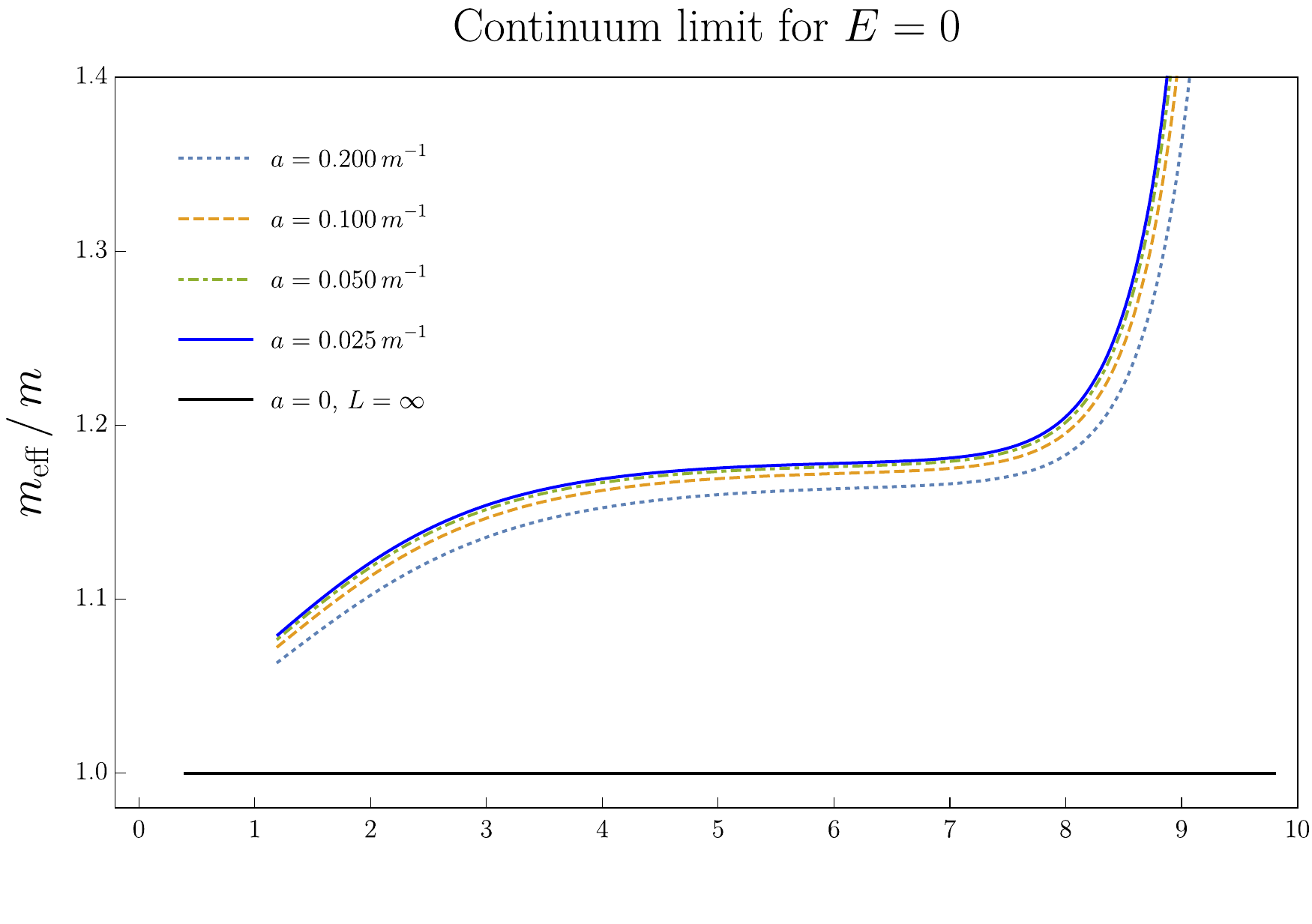} }&
    \parbox[c]{8.9cm}{\includegraphics[width=\linewidth,trim=-.8cm 0 0 0]{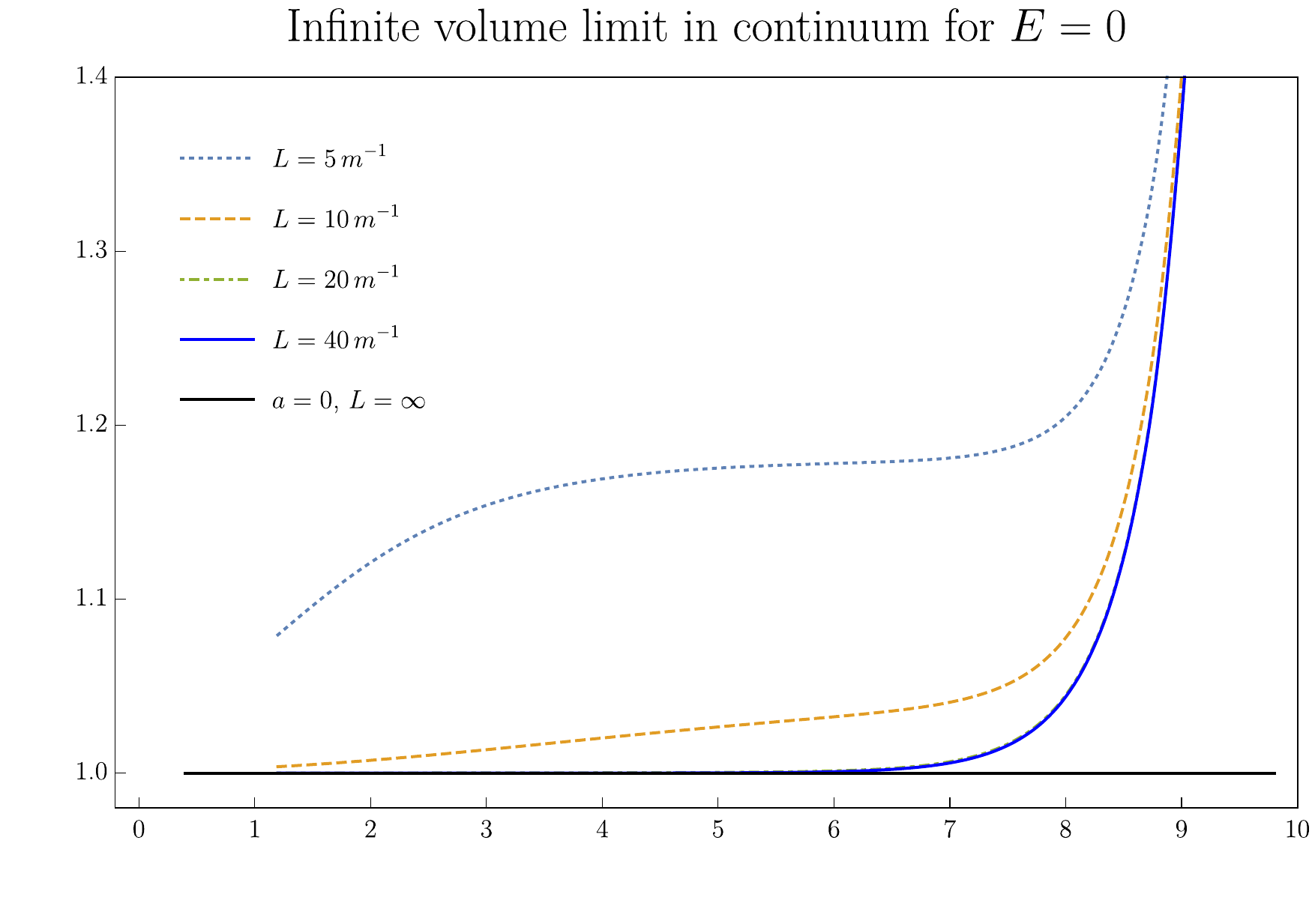} } \\&\\
    \parbox[c]{8.5cm}{\includegraphics[width=\linewidth,trim=0cm 0 0 0]{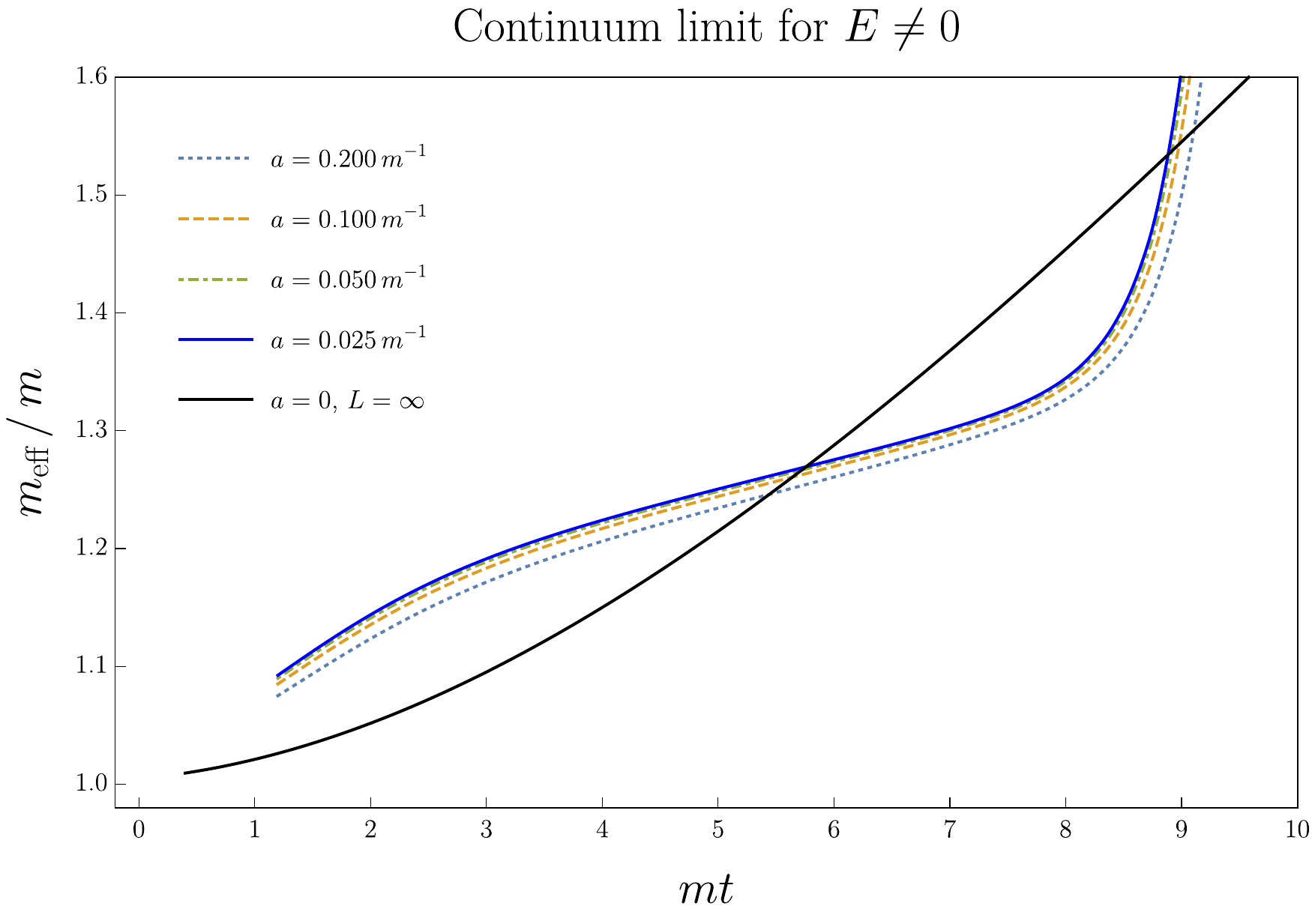} }&
    \parbox[c]{8.9cm}{\includegraphics[width=\linewidth,trim=-.8cm 0 0 0]{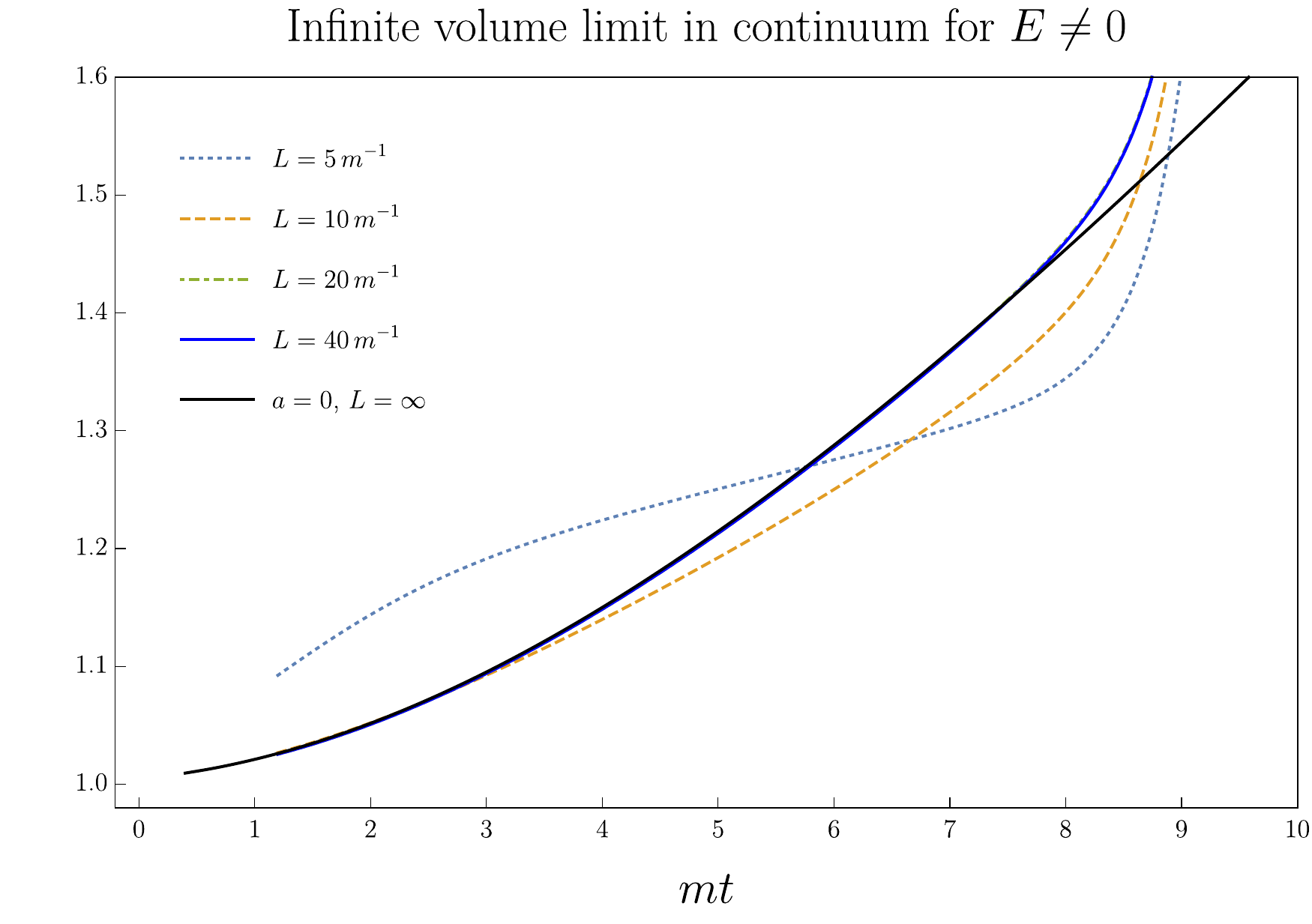} }
    \end{tabular}
    \caption{Comparison of effective mass function extracted from the effective correlator and the infinite-volume one in Eq.~\eqref{eq:infvolcorr}. For plots on the left we study the continuum limit: we make the lattice spacing finer and finer at fixed box size until convergence. For plots on the right, we focus on the finite volume effects. For the lower plots we use a value of the electric field of $a^2q{\cal E}=0.005$. For all plots the effective correlator is computed using Dirichlet boundary conditions in both directions. The lattice size in the time direction is $L_t=10 m^{-1}$ and the source for the propagator is at $t_0=m^{-1}$. }
    \label{fig:model}
\end{figure*}

To extract polarizabilities, we fit the lattice QCD correlator for pions to the correlator for a relativistic scalar particle in two-dimensions. As we discuss later in the paper we use Dirichlet boundary conditions in the direction of the electric field. For transverse directions we use periodic boundary conditions and we expect that the low energy pion states have zero momentum in the transverse directions. As such calculating the particle propagator in $3+1$ dimensions reduces to a two-dimensional problem.
We start with the continuum action in Euclidean space for a massive scalar particle
\beq
  S_E =  \int dt\,dx \left[ \partial_\mu \phi^* \, \partial_\mu \phi + m^2\phi^* \phi \right]
\eeq
where $\phi$ is a complex scalar field.
A background electromagnetic field  can be introduced via a vector potential $A_\mu$ through minimal coupling to the charge,
\beq
  S_E = \int dt\,dx\left[ \left( \partial_\mu +iqA_\mu \right)\phi^* \left( \partial_\mu -iqA_\mu \right) \phi + m^2\phi^* \phi \right].
\eeq
Note that here the charge corresponds to the pion charge $q=e$.
Our effective model is a two-dimensional, discretized version of the action given by~\cite{Rothe:1992nt}
%
%
%
\beq
S_E =\sum_{n,m} \phi_n^* K_{nm} \phi_{m}
\eeq
with
\beqs
K&_{nm} = [4+(am)^2] \delta_{nm} -\\
&- \sum_{\hat{\mu} > 0} \Big[
\delta_{n+\hat{\mu},m} \,e^{-iqaA_\mu(n)} + \delta_{n-\hat{\mu},m}\,e^{+iqaA_\mu(m)}
  \Big] 
\eeqs
where $n$ denotes the sites on the lattice, $a$ is the lattice spacing, and $\hat{\mu}$ accounts for the nearest neighbors of every site on the lattice, \textit{i.e.} $\pm \, \hat{x}$ and $\pm \, \hat{t}$. The two-point correlation function is the inverse of the $K$-matrix:
\beq
\av{\phi_n \phi_m^*} = \frac12 K_{nm}^{-1}.
\label{eq:effprop}
\eeq
This effective correlator can be compared to its infinite-volume, continuum version which can be derived using Schwinger's proper time method~\cite{Schwinger:1951nm,Tiburzi:2008ma}
\begin{equation}
	C(t) = \frac{1}{2} \int_0^\infty ds \, \sqrt{\frac{q\mathcal{E}}{2\pi \sinh(q\mathcal{E}s)}} \,\, e^{-\frac{1}{2} m^2 s - \frac{1}{2} q\mathcal{E} t^2 \coth (q\mathcal{E}s) } 
\label{eq:infvolcorr}.
\end{equation}
The comparison is shown in Fig.~\ref{fig:model} for the effective mass function 
\beq 
m_{\text{eff}}(t)=-\ln{\frac{C(t)}{ C(t-1)}},
\eeq
where the continuum and infinite volume limits are taken. The effective model correlator is computed using Dirichlet boundary conditions. We see that our effective model converges to the correct infinite-volume, continuum version, with or without background electric field.
The effective mass in the presence of the field is not constant even at large times. In fact, it grows as a function of time, probably due to the acceleration of the particle in the field. As we advertised earlier, a different fitting procedure must be developed for the new functional form. 
We also note large differences between the finite-volume and infinite-volume forms for values of $m L\approx 5$ encountered in typical lattice QCD simulations (in this study we focus on the pion and $m$ is the pion mass). Since we are relying on tiny mass shifts to extract polarizabilities, such differences can have a large impact on the results. We will develop in this work a fitting method incorporating the full content of the effective correlator and test it out on a number of lattice QCD ensembles.

\subsection{Boundary conditions}

In a finite volume the background field method requires 
careful handling of the external field at the boundaries. 
Even though the electric field is constant, the electrostatic 
potential $A_\mu$ has discontinuities at the edges of the box.
For real electric fields, these discontinuities cannot be avoided.
One workaround is to use imaginary fields with quantized values
such that the ratio $q{\cal E}/ (2\pi L_x L_t)$ is an integer.
This ensures uniform electric flux through all the $xt$ plaquettes
and while the electrostatic potential is still discontinuous.
One problem with this approach is that there is no obvious analytical
continuation that connects this setup to a situation involving 
physical electric fields, so it is unclear whether the results 
computed in this setup are directly relevant for polarizability.
Another problem is that in this setup different equivalent 
gauge choices for the external fields lead to results that
are not connected by a gauge transformation. For example, in a gauge 
similar to the one used in Ref.~\cite{Detmold:2010ts}, of the form corresponding to Eq.~\eqref{eq:gauge},  the Polyakov loops in the spatial directions depend on the choice of $t_0$ and, since the Polyakov loops are invariant under gauge transformations, different $t_0$ choices could lead to different results.
 This might not be a problem if the two gauges produce the same results, but it has to be verified, {\it a posteriori}.
We choose to work with Dirichlet boundary conditions in both $x$ and $t$ directions. 
In this setup gauge invariance in the external field is guaranteed as we will discuss later.
We can use either real or imaginary electric fields, and non-quantized values for the field strength, as small as needed for the quadratic terms due to polarizability to dominate.

On the other hand, using Dirichlet boundaries introduces a number of 
issues that must be dealt with. First issue is that the lowest energy 
state corresponds to a moving hadron. This is the main reason for 
the discrepancy present in the upper-right panel of Fig.~\ref{fig:am} 
between the effective mass extracted from the finite-volume 
correlator and the infinite-volume one. For neutral hadrons, when 
extracting the energy from exponential decay of the correlators, we 
need an estimate of the hadron momentum to correct for the effect on 
the mass shift~\cite{Lujan:2016ffj}. When fitting the lattice QCD 
correlator to the one from the effective model, this ajustment is done
automatically and the value extracted is directly the rest mass of 
the particle. 

Another important effect is due to the pion interactions with the Dirichlet walls. Since hadrons are extended objects, they interact with the walls differently than the point-like particles described by the effective model. Away from the walls the hadrons move freely and the center-of-mass wavefunction corresponds to a free particle wave, but this does not hold true near the walls. The net effect is that the free wavefunction for the hadrons away from the walls has nodes that are not exactly aligned with the walls. To account for this difference we use a different distance between the walls in the effective model. The difference is related to the {\em reflection scattering length} introduced to capture the effect of this interaction for extended objects~\cite{Lee:2010km,Pine:2012zv}.
This effect will be studied using the pion correlator in the absence of an external field. 

\section{\label{sec:fittingprodecure}Technical details and results}

To fit the effective model we need to construct the hadron wavefunction profile between the Dirichlet walls. This requires us to calculate lattice QCD correlators as a function of both $x$ and $t$, the directions where we use Dirichlet boundary conditions. In the $y$ and $z$ directions we use periodic boundary conditions and we project the correlator to zero momentum in these directions,
\beq
G(x,t; A_\mu, L)=\sum_{y,z} \langle 0|\hat{O}(x,y,z,t)\hat{O}^\dagger(x_s,y_s,z_s,t_s)|0\rangle,
\eeq
where $\hat{O}=\bar{d}\gamma_5 u$ is the interpolator for $\pi^+$ in the two-point function, is $(x_s,y_s,z_s,t_s)$ is the source for the quark correlators. Note that due to isospin symmetry between the $u$ and $d$ quarks,  $\pi^+$ and  $\pi^-$ ($\hat{O}=\bar{u}\gamma_5 d$) have identical polarizability.

For the effective model we use the propagator derived in Eq.~\eqref{eq:effprop}
\beq
G_0(x,t; A_\mu, \tilde L) = K_{(x,t); (x,t)_s}^{-1} \,.
\eeq
Above we emphasized that the correlators are functions of the external field $A_\mu$ and 
the distance between the walls: $\tilde L$ for the effective correlator and $L$ for QCD calculation.

We have generated the correlators with the position of quark sources fixed at $x_s/a = N_x/2+1$ and $t_s/a=6$. 
We use a coordinate system where the Dirichlet walls are positioned at $x=0$ and $x=L$ in the lattice QCD box. 
The effective model is aligned such 
that the source is at the same position and the walls are at $x=(L-\tilde L)/2$ and $x=(L+\tilde L)/2$.
In effect the setup keeps the source position the same and changes the distance to the $x$-direction walls by the same amount.
For both QCD and the effective model the time boundaries are  $t=0$, and $t/a=(N_t+1)$. 
For QCD correlator, we take advantage of translation invariance of the system in $y$ and $z$ directions, and use 
sources at 64 locations for each configuration to obtain more statistics.
The sources are spaced regularly at 8 positions on a $y$-$z$ grid and we use 8 different 
time translations. Note that since the gauge fields are generated with (anti-)periodic boundary
conditions in time, time-translation is a symmetry of the ensemble. We apply the Dirichlet boundary
conditions in time after the translation. A similar symmetry exists also in the $x$-direction but we
did not take advantage of it in this study.

To find the appropriate distance between the walls for the model,
we fit the lattice correlator to the effective model using different
sizes for the model. We fit the correlators away from 
the walls to insure that the difference in the interactions with 
the walls do not affect the fit. The fit ranges, in both $x$- and
$t$-directions are reported in Table~\ref{tab:ens}.

\begin{table}[b]
\setlength{\tabcolsep}{3.5pt}
\renewcommand{\arraystretch}{1.4}
	\centering
	\begin{tabular*}{0.99\columnwidth}{@{\extracolsep{\stretch{1}}}cccccc@{}}
		\toprule
		Ensemble & $N_x N_y N_z N_t$ & $N_{\texttt{cfg}}$ & $\tilde L/a$ & $x_i - x_f$  & $t_i - t_f$\\ 
		\midrule
		EN1&$16 \times 16^2 \times 32$ & 450 & 14.00(13)& $07-10$ & $15-24$ \\
		EN2&$24 \times 24^2 \times 48$ & 300 & 21.50(17)& $07-18$ & $15-32$ \\
		EN3&$30 \times 24^2 \times 48$ & 300 & 26.50(18)& $07-24$ & $15-32$ \\
		EN4&$48 \times 24^2 \times 48$ & 300 & 44.00(72) & $07-42$ & $15-32$ \\
		\bottomrule
	\end{tabular*}
	\caption{Dynamical QCD ensembles on elongated lattice used in our study. In all cases $a = 0.1245\, \fm$ and $m_\pi = 315\, \MeV$. The last three columns show the effective spatial size and fit ranges.
	}
	\label{tab:ens}
\end{table}
\begin{figure*}[t]
    \centering
    \begin{tabular}{cc}
    \parbox[c]{8.5cm}{\includegraphics[width=\linewidth,trim=0cm 0 0 0]{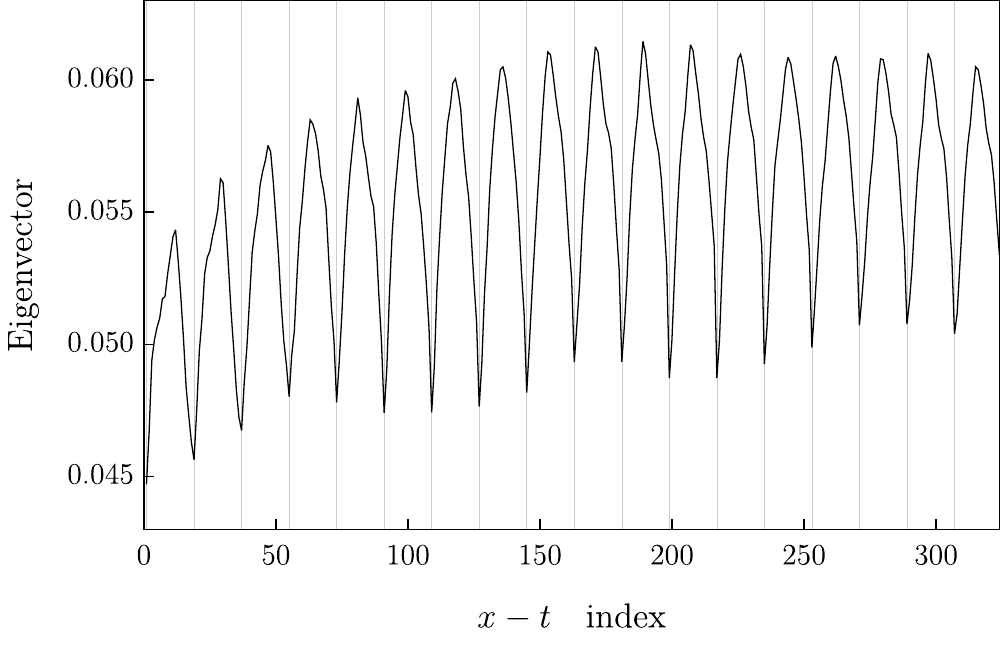} } &
    \parbox[c]{8.5cm}{\includegraphics[width=\linewidth,trim=0cm 0 0 0]{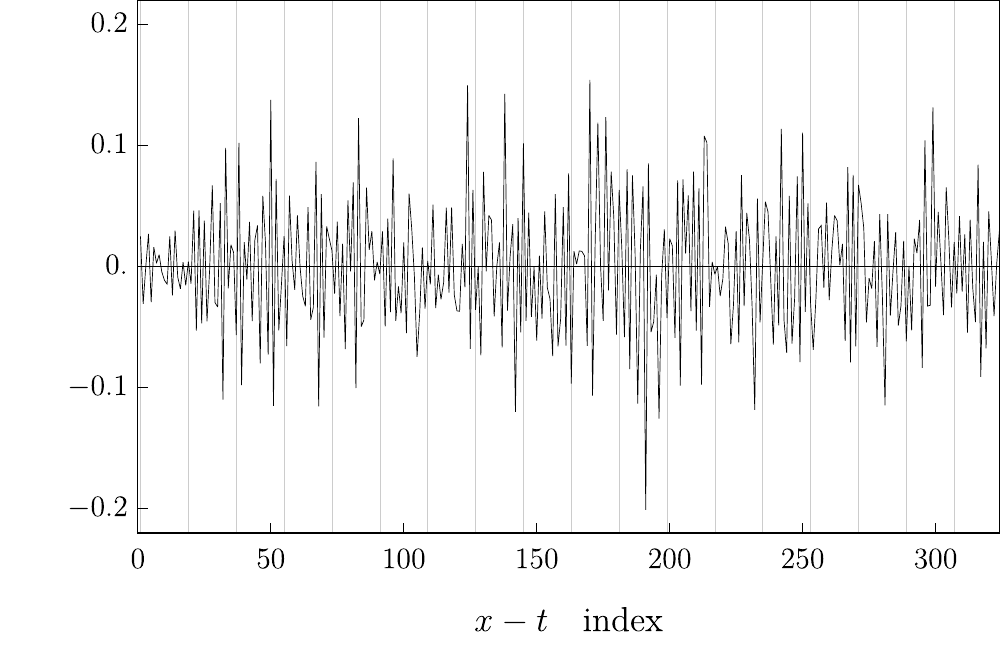} }
    \end{tabular}
    \caption{Eigenvectors corresponding to the largest (left) and smallest (right) eigenvalue of the covariance matrix, 
    when including all points in the range listed in Table~\protect\ref{tab:ens} for EN3. 
    The thin vertical lines are the lines separating different time-slices; within each time slice, 
    $x$ changes from left to right from $x_i$ to $x_f$. 
    }
    \label{fig:highfreq}
\end{figure*}

\subsection{Fitting procedure}

To extract the mass from the lattice correlator we determine the mass in the effective model
that minimizes the distance between the QCD correlator $G$ and the effective one $G_0$. The
distance between the two correlators is defined via a $\chi^2$ function. The design of this
distance function was guided by two main concerns: the fit procedure should produces the same results 
for all possible equivalent external field gauges and we want to accommodate the fact that the
correlator is a complex valued function.

Assume that we perform on each configuration the measurements $y_i$ that are complex valued. The
effective model value that is designed to match $y_i$ is assumed to be $f_i$. We define the $\chi^2$-function
\beq
\chi^2 \equiv \delta^\dagger C^{-1} \delta \quad\text{with the {\em residues}}\quad \delta_i \equiv \av{y_i} - f_i \,,
\eeq 
and the covariance matrix
\beq
C_{ij} \equiv \av{(y_i-\av{y_i})(y_j - \av{y_j})^* } \,.
\eeq
The covariance matrix is hermitian and the $\chi^2$-function is real and positive-definite. The minimization
process will vary the parameters of the effective model to minimize $\chi^2$.

We are now interested in the action of the gauge transformations on the distance functions through 
the observables $y_i$. Assuming that the transformation acts linearly on the observables such that 
$y' = Ty$~(that is $y'_i=T_{ij}y_j$), with $T$ the same matrix for all configurations. 
Then $\av{y'}=T\av{y}$ and {\em if} we have $f'=Tf$, then $\delta' = T\delta$. 
Similarly for the covariance matrix we have $C' = TCT^\dagger$ and
the distance function is then invariant, that is $(\chi^2)' = (\delta')^\dagger (C')^{-1}\delta'= \chi^2$.
This condition is satisfied if the observables $y_i$ are any subset of the point-to-point propagator 
function~$G(x,t; A_\mu, L)$ since under a gauge transformation, $(A,\phi)\to(A',\phi')$ with
\beqs
 A'_\mu(x,t) &= A_\mu(x,t)+[\Lambda((x,t)+\hat\mu)-\Lambda(x,t)]/a\\
\phi'(x,t) &= e^{iq\Lambda(x,t)}\phi(x,t) \,,
\eeqs
we have
\beq
G(x,t; A_\mu', L) = e^{iq[\Lambda(x,t)-\Lambda(x_s,t_s)]} G(x,t; A_\mu, L) \,. 
\eeq
Above we assumed that we use imaginary electric field. The same relation holds for the effective
theory propagator~$G_0$ so the distance between the point-to-point propagators is invariant under
gauge transformations.


Given the discussion above it is then tempting to fit the lattice propagator at a subset of the $(x,t)$ positions,
in the range included in Table~\ref{tab:ens}.
The problem is the presence of high frequency modes in the propagators close to the ultraviolet cutoff that are distorted due to
lattice discretization. These modes can have an important contribution to the $\chi^2$-function. The smallest eigenvalues of 
the covariance matrix $C$ will have the biggest contribution to the $\chi^2$. This means that the lowest eigenvalues will drive
the fitter and force the minimizer to a point in the parameter space that matches the distorted high frequency modes in the 
lattice QCD data instead of the physical long-range two-point function we are interested in. This is indeed the case for our
calculation. In Fig.~\ref{fig:highfreq} we show the modes corresponding to the lowest and the highest eigenvalue in the
covariance matrix, when we take the observables to be all space-time points in the range indicated in Table~\ref{tab:ens} for 
ensemble EN3. We see that the most important modes for the fitter, the ones that correspond to the lowest eigenvalue of the
correlation matrix, are the ones that fluctuate rapidly along the spatial directions, the
distorted high-frequency modes.
To remedy this problem, we need to filter out the these high frequency modes from our observables.

\begin{figure*}[htb!]
    \centering
    \begin{tabular}{cc}
    \parbox[c]{8.5cm}{\includegraphics[width=\linewidth,trim=0cm 0 0 0]{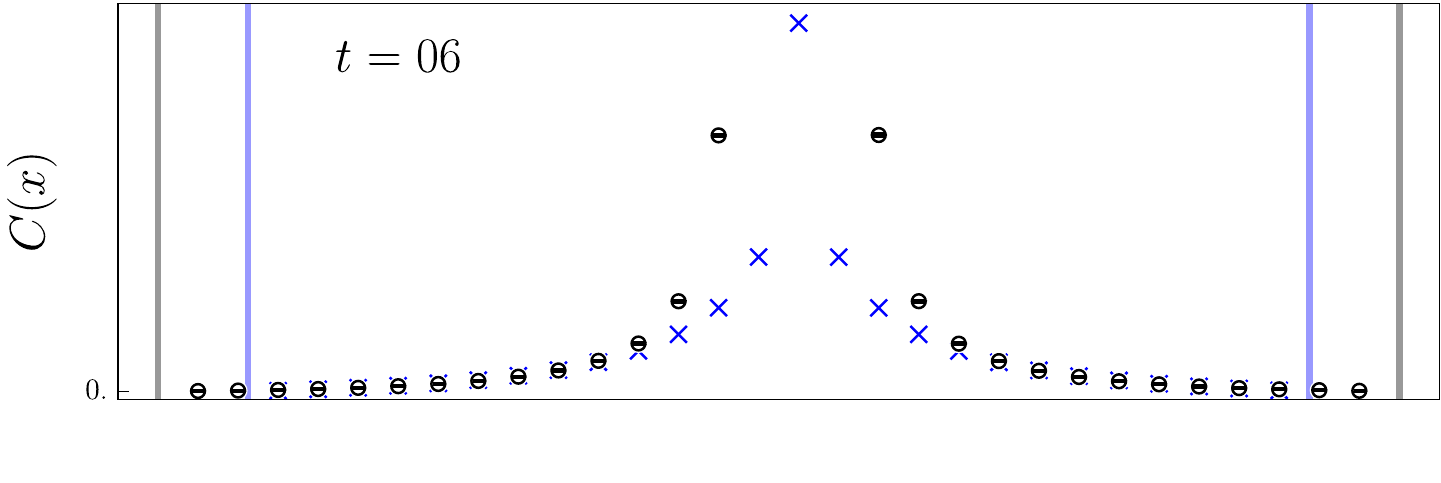} }& \vspace{-.8cm} \hspace{-.6cm}
    \parbox[c]{8.9cm}{\includegraphics[width=\linewidth,trim=-.9cm 0 0 0]{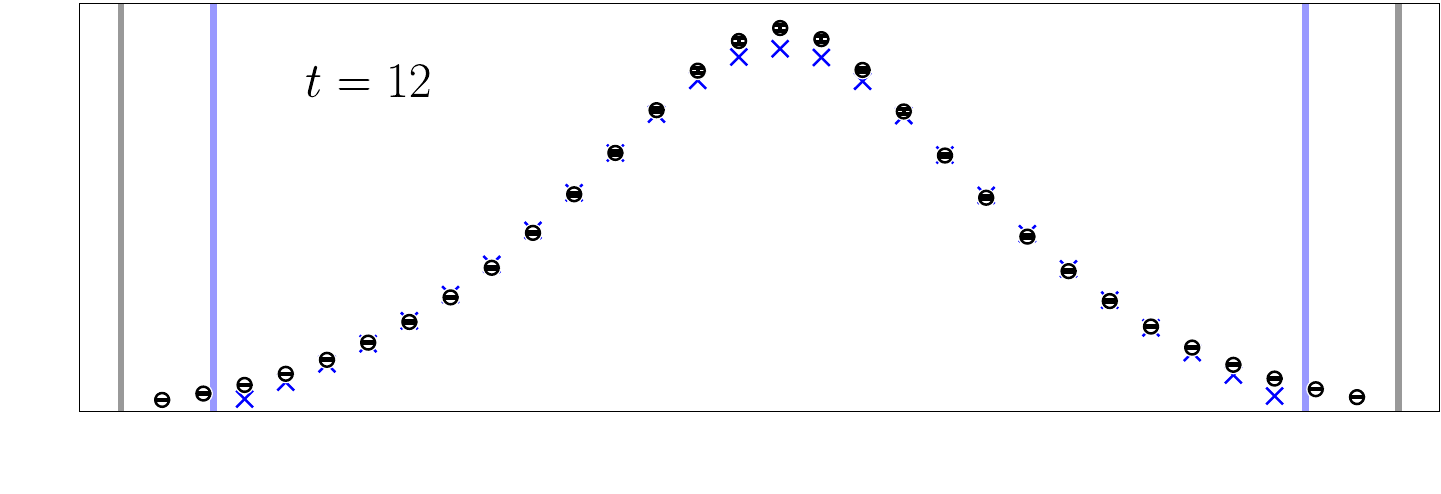} } \\&\\
    \parbox[c]{8.5cm}{\includegraphics[width=\linewidth,trim=0cm 0 0 0]{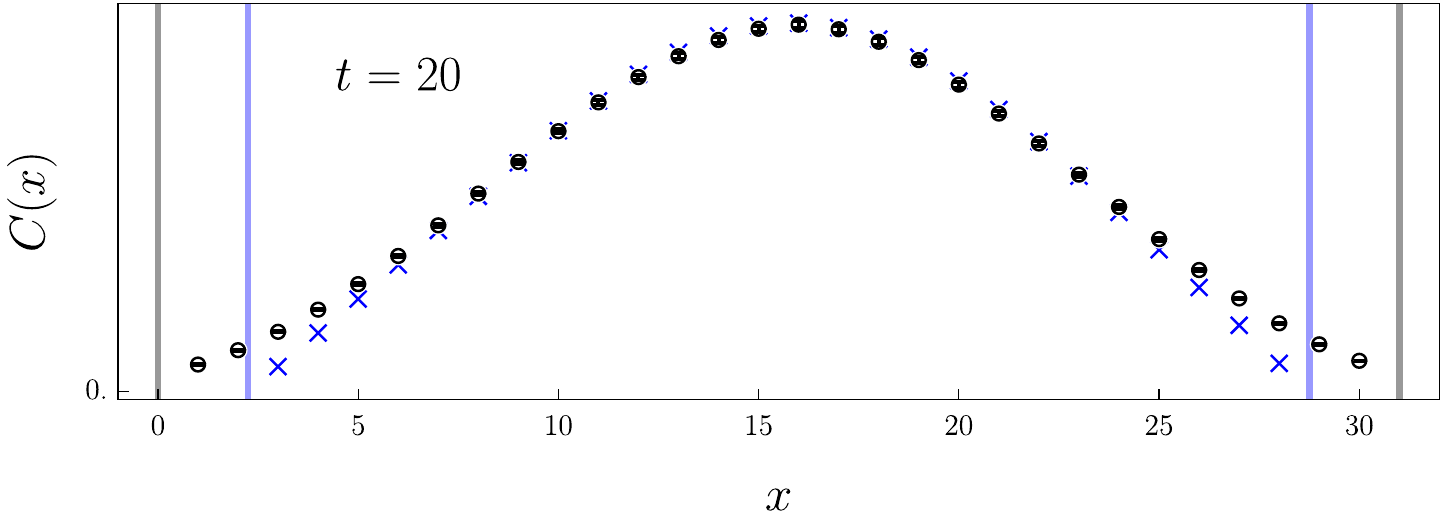} }& \hspace{-.5cm}
    \parbox[c]{8.9cm}{\includegraphics[width=\linewidth,trim=-.9cm 0 0 0]{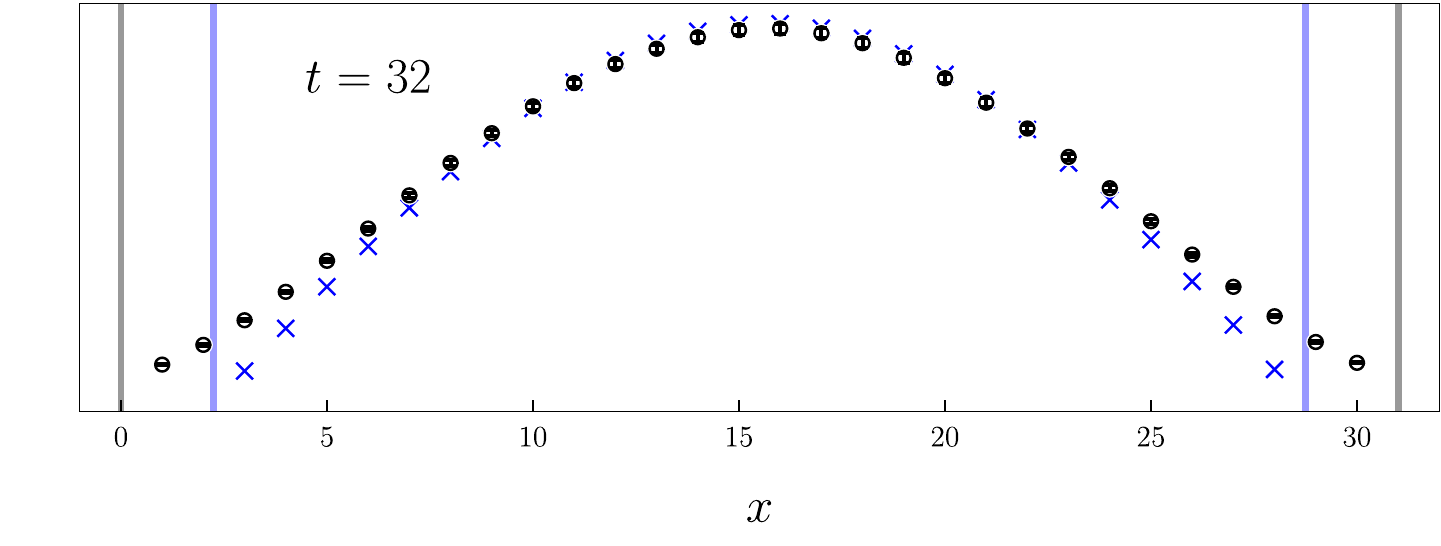} }
    \end{tabular}
    \caption{Spatial profile of the lattice correlator $G(x,t)$ (black) in zero electric field for EN3 is compared to the effective one (blue) at fixed time slices. The corresponding Dirichlet boundaries are represented by vertical lines.}
    \label{fig:Zprop}
\end{figure*}
\begin{figure*}[htb!]
    \centering
    \begin{tabular}{cc}
    \parbox[c]{8.7cm}{\includegraphics[width=\linewidth,trim=0cm 0 0 0]{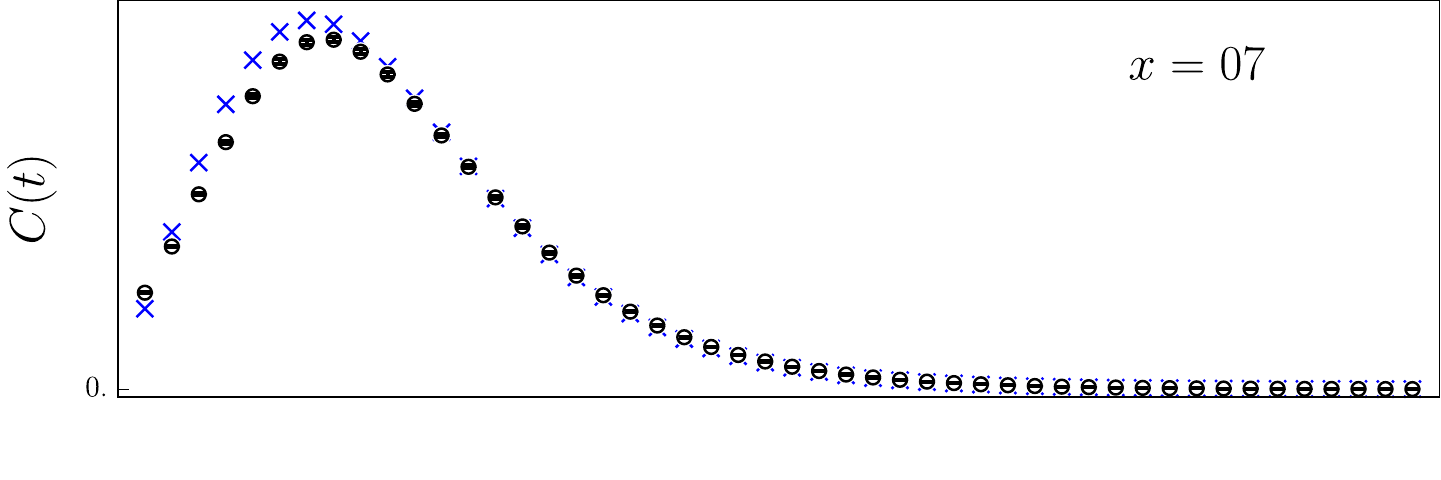} }& \vspace{-.75cm} \hspace{-.95cm}
    \parbox[c]{8.9cm}{\includegraphics[width=\linewidth,trim=-.9cm 0 0 0]{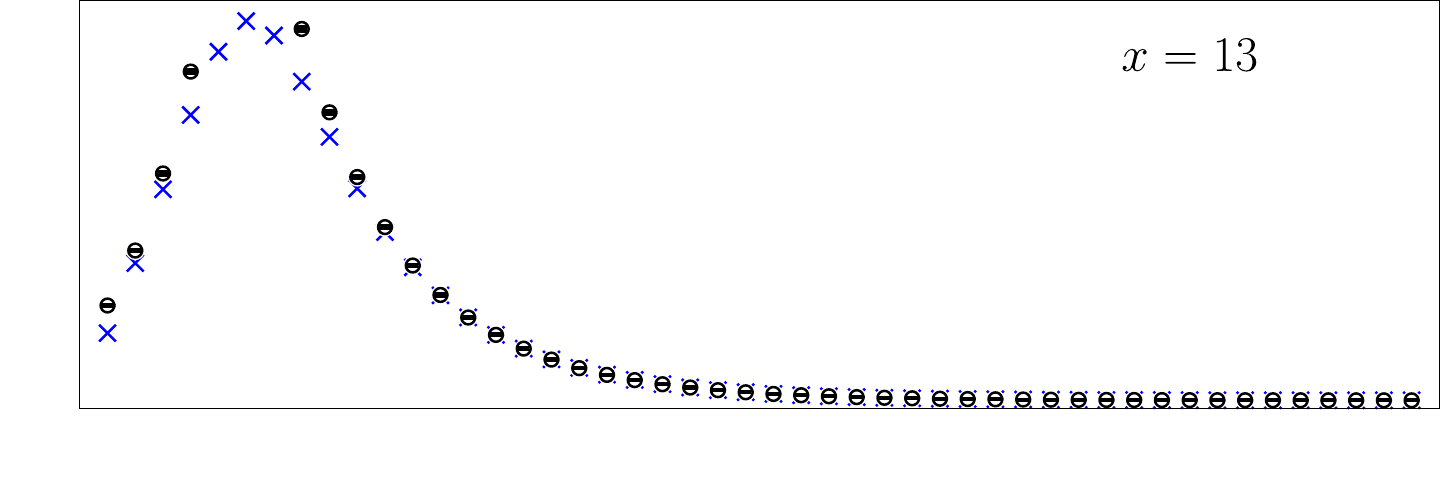} } \\&\\
    \parbox[c]{8.7cm}{\includegraphics[width=\linewidth,trim=0cm 0 0 0]{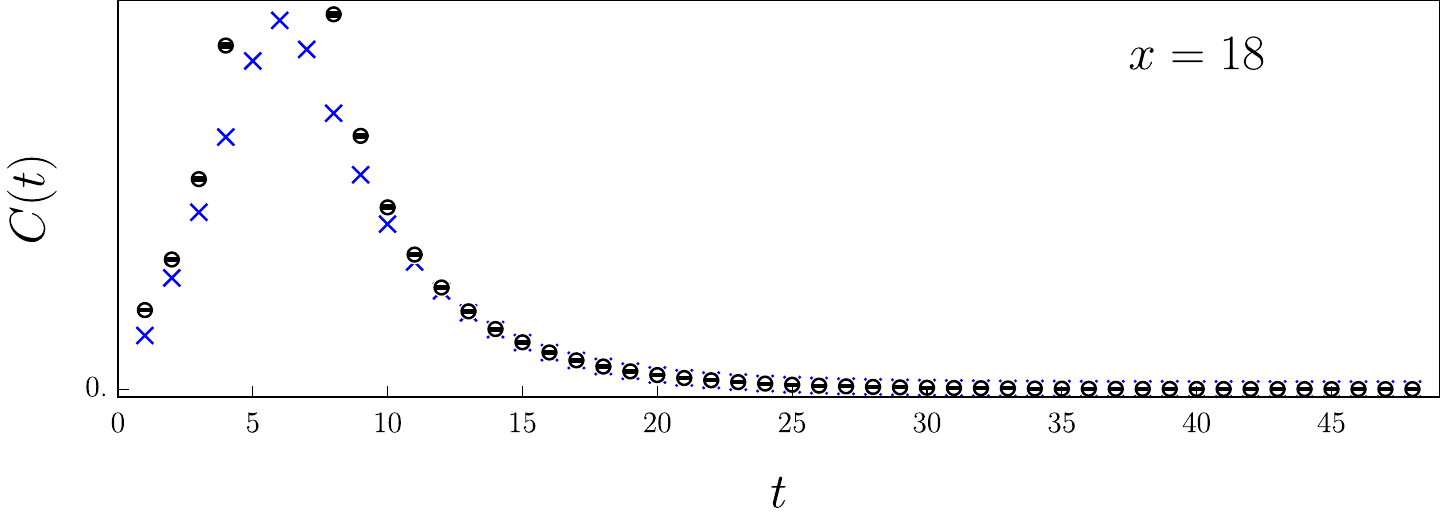} }& \hspace{-.85cm}
    \parbox[c]{8.9cm}{\includegraphics[width=\linewidth,trim=-.9cm 0 0 0]{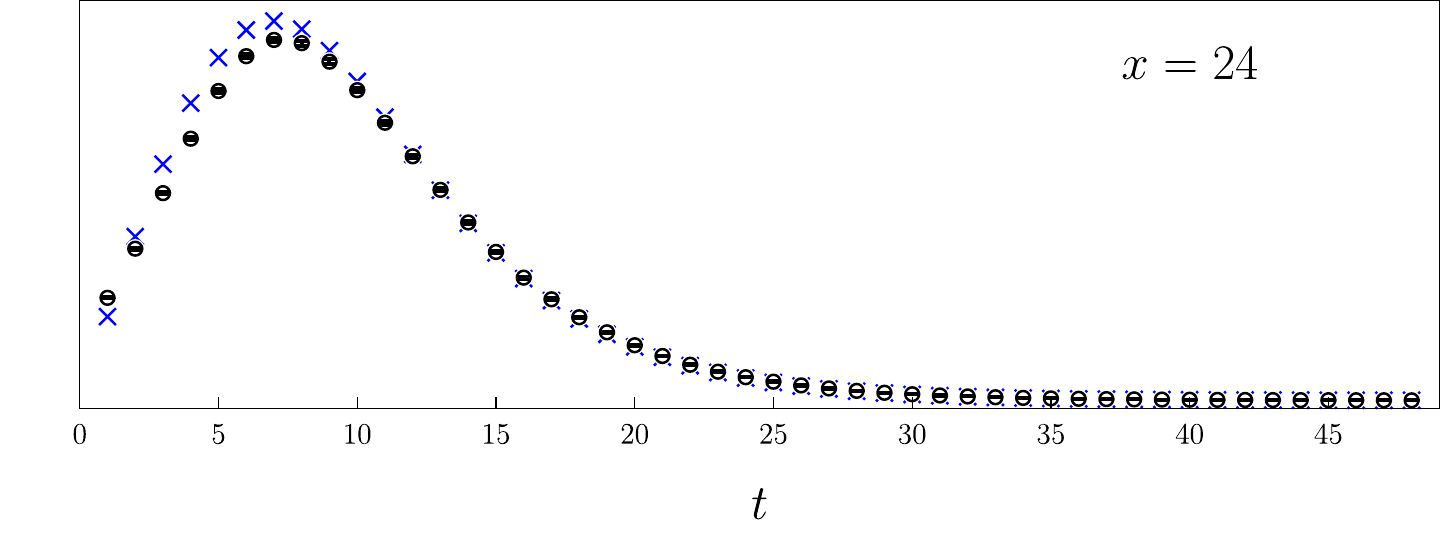} }
    \end{tabular}
    \caption{Similar to Fig~\protect\ref{fig:Zprop}, but for amplitude decay of the correlator at fixed $x$ locations.}
    \label{fig:xprofile}
\end{figure*}

For the zero-field propagators, we can just sum over the spatial position to filter out the high-frequency modes.
This is akin to zero momentum projection used when fitting propagators with periodic boundary conditions. 
For our case this is only used to reduce the influence of the high-frequency modes, not to project to a particular 
momentum state. Our summation only extends over the {\em interior} points, away from the Dirichlet walls.  
However, this does not work for non-zero field propagators since gauge transformations act non-linearly on
the observable defined via this summation. A different form is required to preserve gauge-invariance.
In this work we use the following observables:
\beqs
  y(t) &= \sum_{x=x_i}^{x_f} L(x,t) G(x,t; A_\mu=0, L ) \,\\
  f(t) &= {\cal A} \sum_{x=x_i}^{x_f} L(x,t) G_0(x,t; A_\mu=0, {\tilde L} ) \,.
  \label{eq:proj}
\eeqs
where 
\beq
L(x,t) = \prod_{x'=x_i}^{x-a} e^{-iqaA_\mu(x',t)} \,,
\eeq
and we fit for the amplitude ${\cal A}$ and mass $am$ that enters the definition of $G_0$. 
For the zero field case $L(x,t)=1$ and the summation produces the usual filtering. For the
non-zero case, the {\em gauge line} $L(x,t)$ transforms as $L'(x,t) = e^{iq[\Lambda(x_i,t)-\Lambda(x,t)]} L(x,t)$.
The observables then change under gauge transformation as 
\beq
y'(t) = e^{iqa[\Lambda(x_i,t)-\Lambda(x_s,t_s)]} y(t)
\eeq
and similarly for $f(t)$, and the distance function becomes gauge invariant.

The energy shift induced by the background field is very small, much smaller than
the stochastic error on the extracted hadron masses. To extract the shift reliably
we need to take into account the correlations between zero-field and non-zero field correlators.
We perform simultaneous fits of the zero-field and non-zero field correlators.
The {\em residue} vector $\delta = \{\delta_0, \delta_{\cal E}\}$ includes the zero-field residue $\delta_0$ and the
residue in the presence of the field $\delta_{\cal E}$ defined using the fit form
\beq
  f(t)= ({\cal A}+\Delta {\cal A}) \sum_{x=x_i}^{x_f} L(x) G_0(x,t; A_\mu, {\tilde L},am+\Delta am ) \,.
  \label{eq:fit}
\eeq
 The enlarged covariance matrix  can be written as
\beq
C=\left(
\begin{array}{cc}
C_{00} & C_{0{\cal E}} \\
C_{{\cal E} 0} & C_{{\cal E}{\cal E}} 
\end{array}
\right ).
\eeq
The enlarged $\chi^2$ remains gauge invariant.
Above we made explicit the dependence on the mass parameter. The fit parameters are
${\cal A}$, $am$, $\Delta {\cal A}$, and $\Delta am$. The mass shift $\Delta am$ is used to compute
the polarizability.

\begin{figure*}
    \centering
    \begin{tabular}{cc}
    \parbox[c]{8.5cm}{\includegraphics[width=\linewidth,trim=0cm 0 0 0]{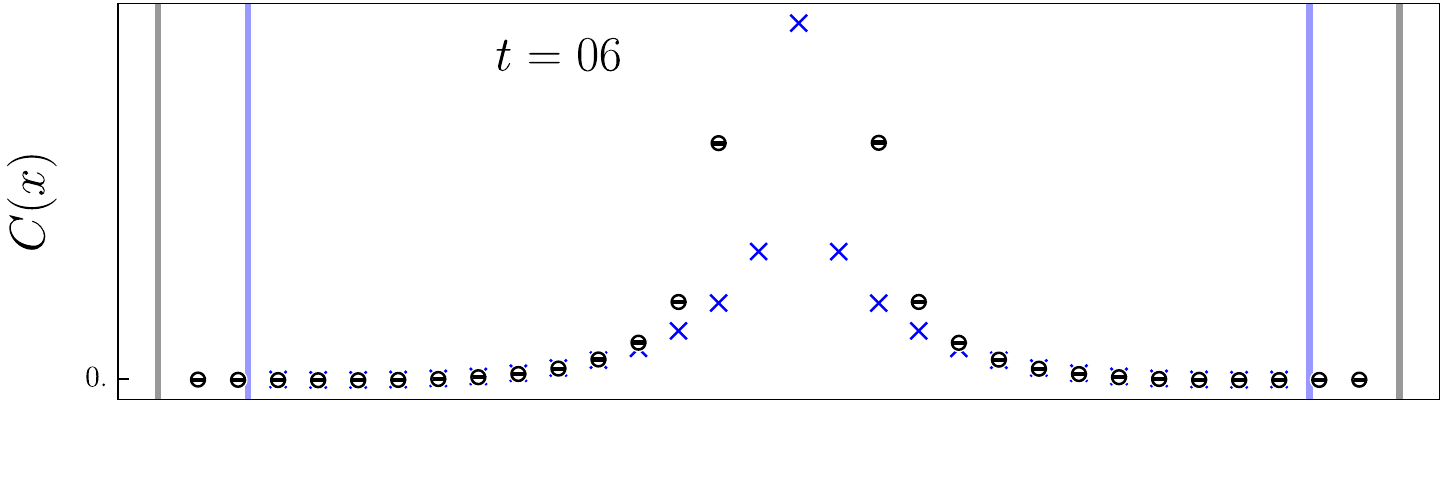} }&  \vspace{-.8cm} \hspace{-.6cm}
    \parbox[c]{8.9cm}{\includegraphics[width=\linewidth,trim=-.9cm 0 0 0]{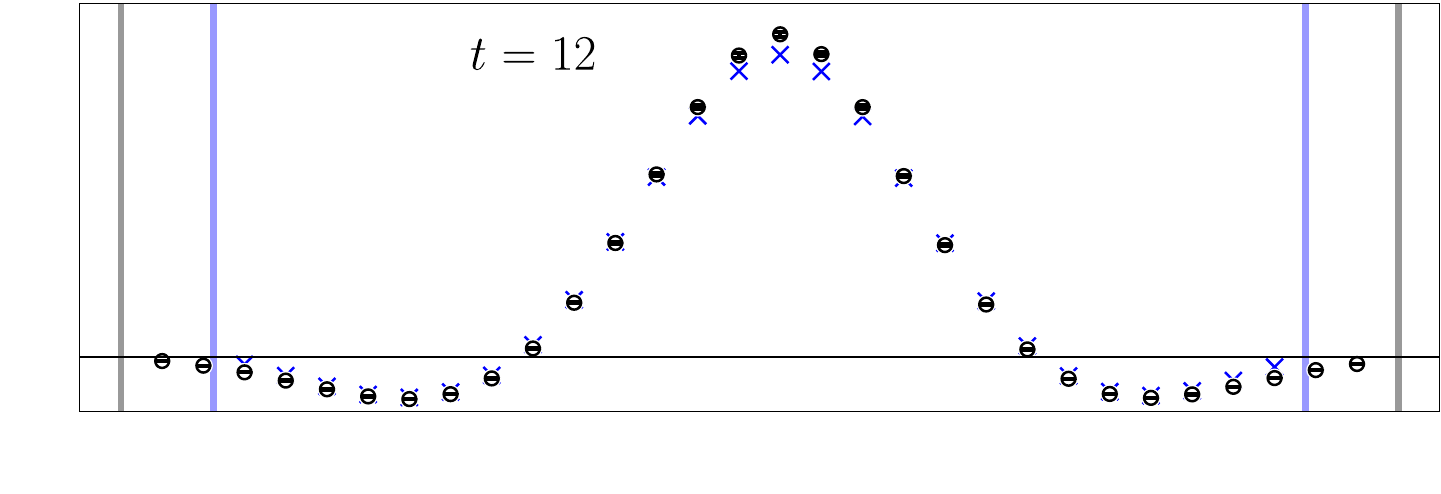} } \\&\\
    \parbox[c]{8.5cm}{\includegraphics[width=\linewidth,trim=0cm 0 0 0]{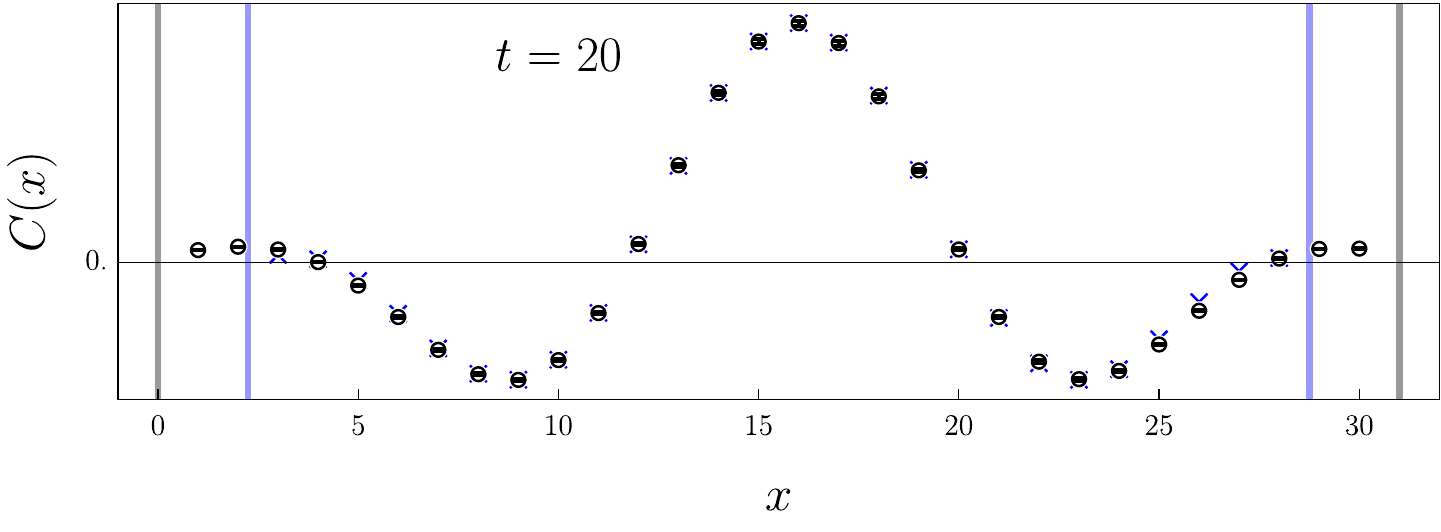} }&  \hspace{-.5cm}
    \parbox[c]{8.9cm}{\includegraphics[width=\linewidth,trim=-.9cm 0 0 0]{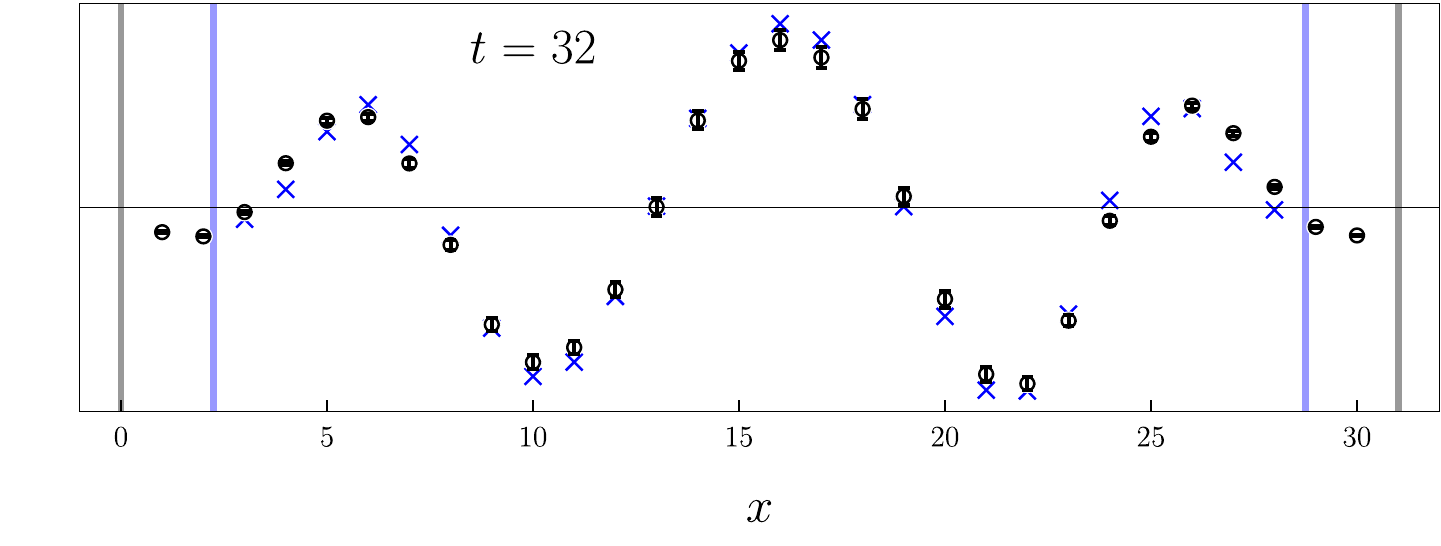} }
    \end{tabular}
    \caption{Real part of the lattice correlator $G(x,t)$ (black) in the presence of the electric field for EN3 is compared to the effective one (blue) at fixed time slices. The corresponding Dirichlet boundaries are represented by vertical lines.}
    \label{fig:Rprop}
\end{figure*}

\begin{figure*}
    \centering
    \begin{tabular}{cc}
    \parbox[c]{8.5cm}{\includegraphics[width=\linewidth,trim=0cm 0 0 0]{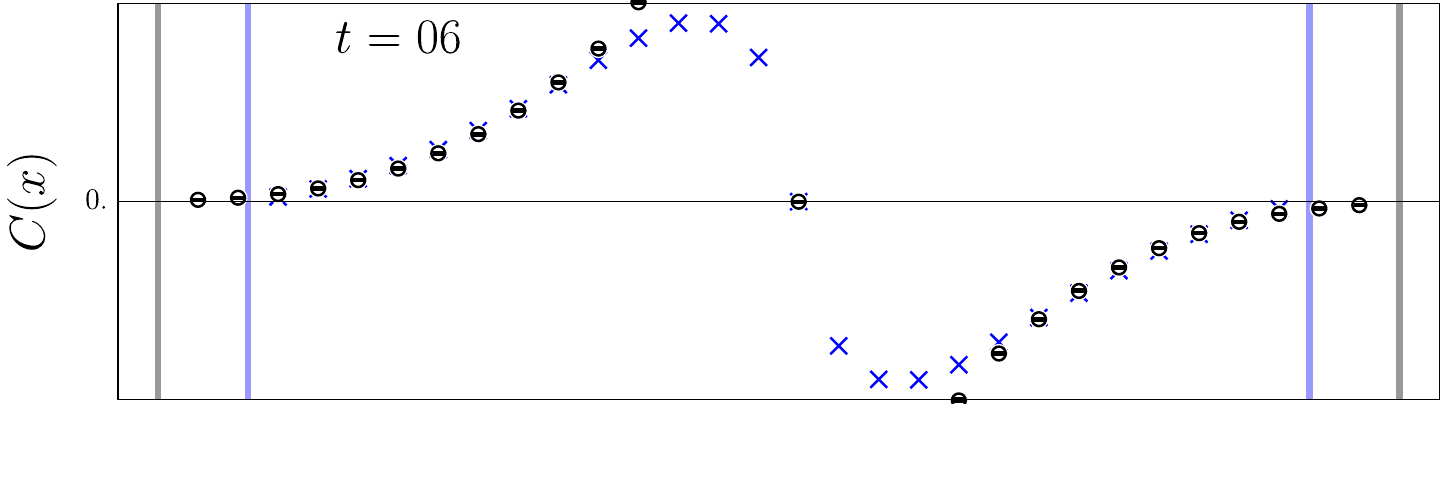} }&  \vspace{-.8cm} \hspace{-.6cm}
    \parbox[c]{8.9cm}{\includegraphics[width=\linewidth,trim=-.9cm 0 0 0]{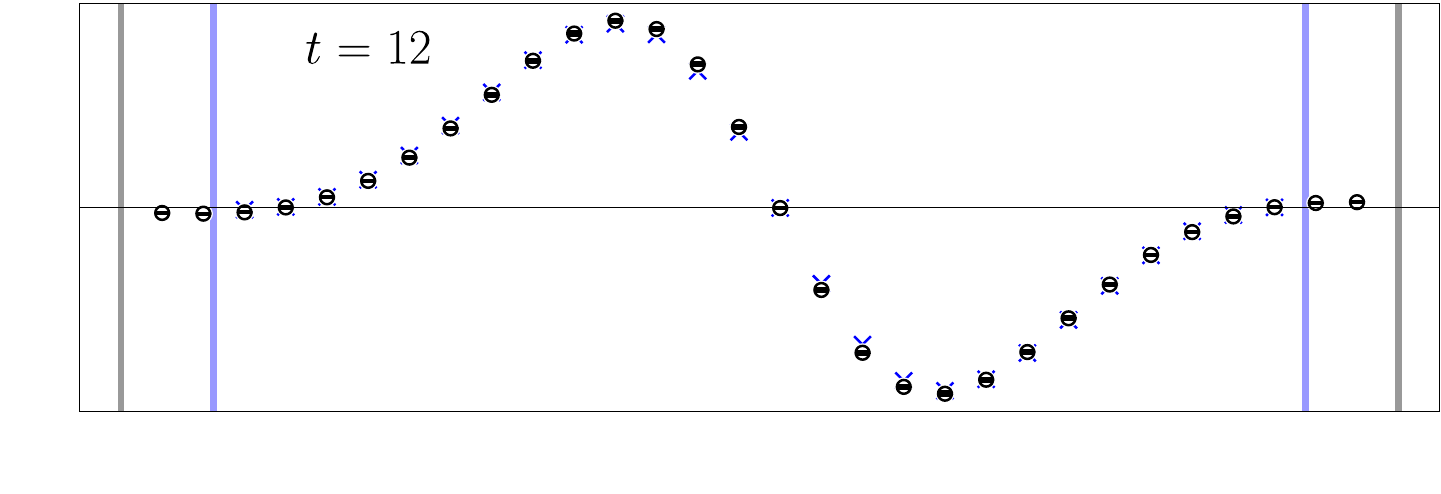} } \\&\\
    \parbox[c]{8.5cm}{\includegraphics[width=\linewidth,trim=0cm 0 0 0]{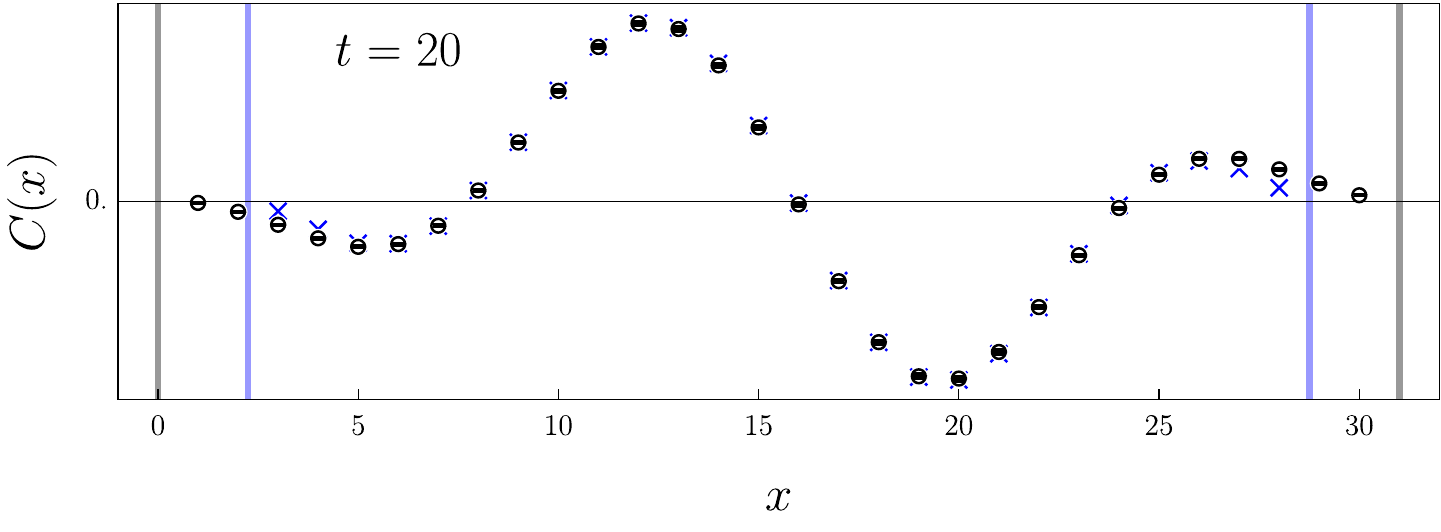} }& \hspace{-.5cm}
    \parbox[c]{8.9cm}{\includegraphics[width=\linewidth,trim=-.9cm 0 0 0]{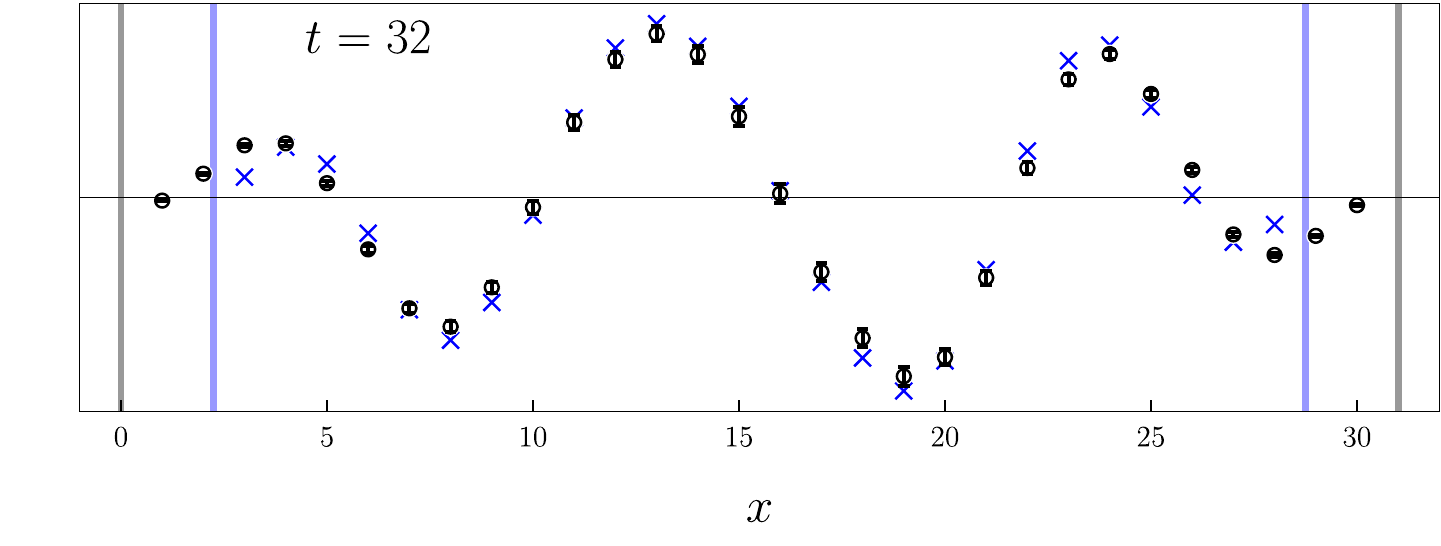} }
    \end{tabular}
    \caption{Similar to Fig~\protect\ref{fig:Rprop}, but for the imaginary part.}
    \label{fig:Iprop}
\end{figure*}

\subsection{Effective range}

As discussed earlier, the {\em effective size} $\tilde L$, the distance between the Dirichlet walls in the
effective model, is slightly different from the distance between the walls in the QCD ensemble due to the
interaction between the hadrons and the walls.
To determine $\tilde L$ we use zero-field lattice QCD correlators. The ensembles used in this study are summarized in Table~\ref{tab:ens}. For details about the generation of the ensembles see~\cite{Niyazi:2020erg}. 

\begin{figure}[t]
    \centering
    \begin{tabular}{c}
    \parbox[c]{0.95\columnwidth}{\includegraphics[width=\linewidth,trim=0cm 0 0 0]{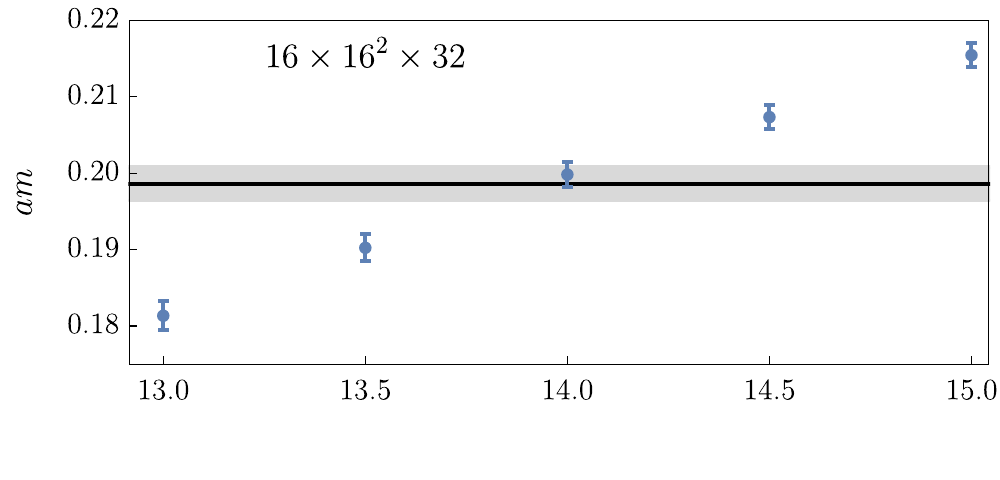} } \vspace{-0.3cm} \\
    \parbox[c]{0.95\columnwidth}{\includegraphics[width=\linewidth,trim=0cm 0 0 0]{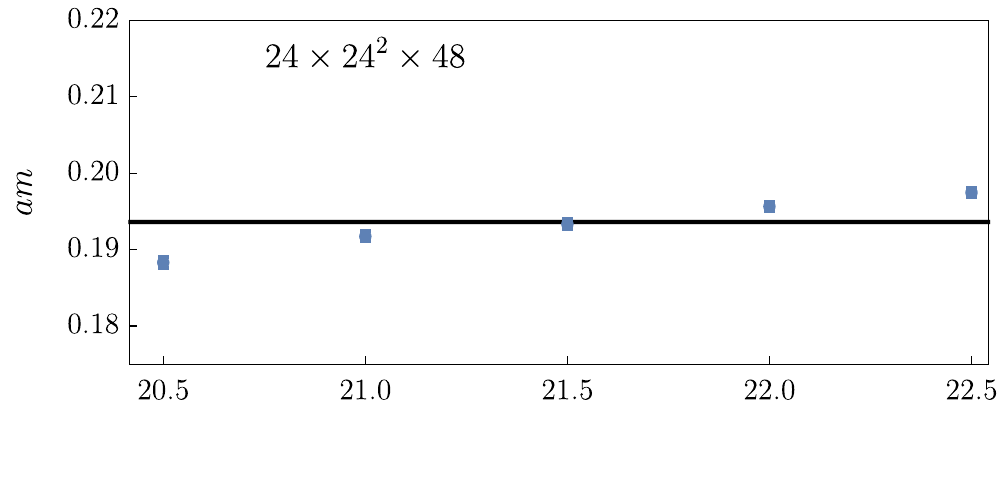} } \vspace{-0.3cm} \\
    \parbox[c]{0.95\columnwidth}{\includegraphics[width=\linewidth,trim=-.0cm 0 0 0]{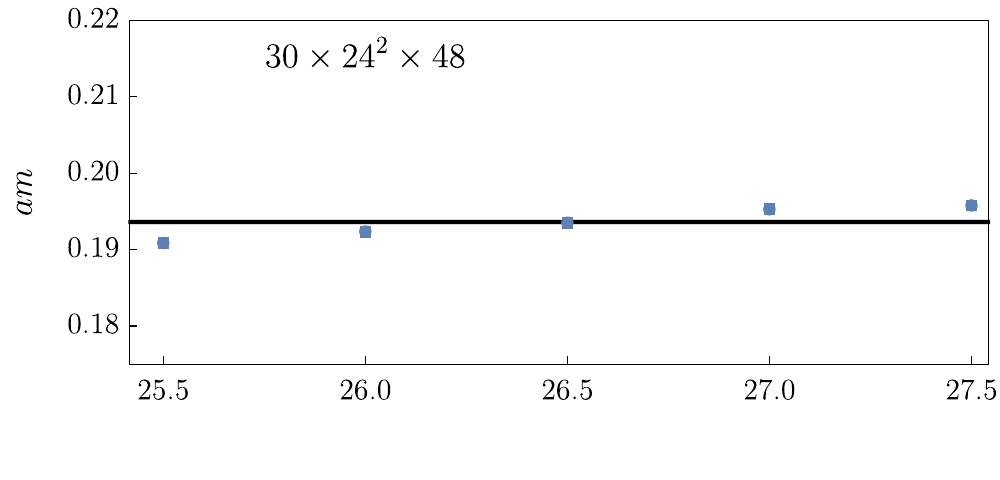} } \\
    \parbox[c]{0.95\columnwidth}{\includegraphics[width=\linewidth,trim=-.0cm 0 0 0]{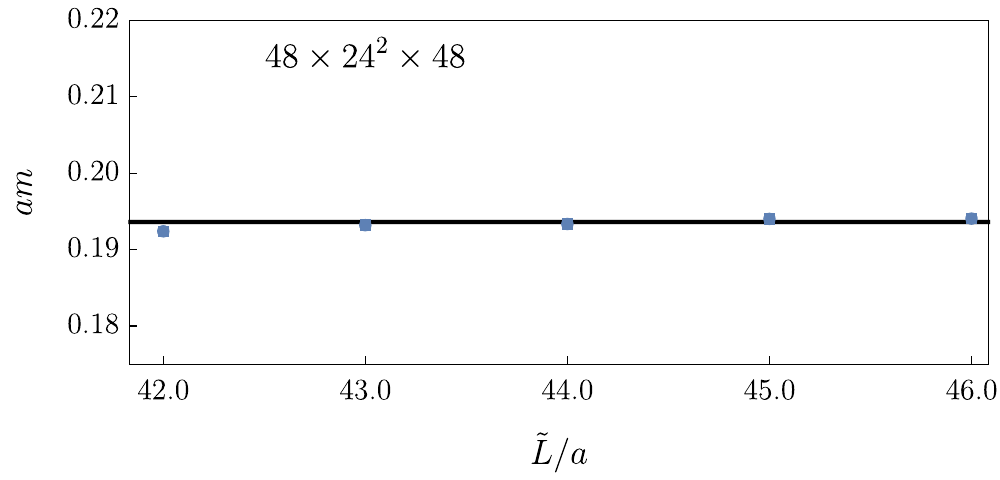} }
    \end{tabular}
    \caption{Dependence of $am$ on the distance between the Dirichlet walls for zero-field case. This is used to determine the effective lattice size $\tilde L$ in Table~\protect\ref{tab:ens}.
    }
    \label{fig:am}
\end{figure}

For each ensemble, we have done multiple fits to map out the dependence of the extracted mass on $\tilde L$. We choose the $x$ and $t$ ranges as wide as possible while maintaining a reasonable $\chi^2$, and look for an effective size $\tilde L$ that reproduces the mass obtained from periodic boundary conditions on the same ensembles~\cite{Culver:2019qtx}. For the three larger ensembles the finite-volume corrections for the pion mass are small and we used $am_\pi =0.1936(2)$. For the smallest ensemble, EN1, we used $am_\pi =0.1986(22)$~\cite{Lujan:2016ffj}. The error-bars on $\tilde L$ were determined by jackknife. The dependence of the extracted mass on $\tilde L$ for our ensembles is shown in Fig.~\ref{fig:am}. We will discuss later the sensitivity of the mass-shift extraction on the effective size.  The $\tilde L$ values determined from our ensembles and the fit ranges used are added to Table~\ref{tab:ens} and  will be used in all of our subsequent fits. Note that the $\tilde L$ values were rounded up to the nearest half lattice-spacing to make the effective model calculations easier.

To see how well the effective model describes the lattice QCD correlators, we show in Fig.~\ref{fig:Zprop} 
some snapshots of the spatial profile (wavefunction) in the absence of electric field. Indeed, the Dirichlet walls 
force a standing wave in the $x$ direction. For the effective model there are no excited states in contrast to QCD where other
excitations are present. However, in the pion channel the excited states lie much higher and the ground state pion becomes dominant
as we move away from the source. We can see from the figure that this is indeed the case: the two correlators match very well as 
time progresses. This suggests that our effective models correctly captures the relevant dynamics. We note here that the correlators
match well only after we adjust the effective distance $\tilde L$ appropriatedly. The same match is observed in the time decay 
of the correlation at fixed $x$ locations, as shown in Fig.~\ref{fig:xprofile}. 


%


In the presence of the background electric field the correlators become complex.
We show in  Fig.~\ref{fig:Rprop} and Fig.~\ref{fig:Iprop} 
the real and imaginary parts separately of the charged particle correlator. 
For the effective correlator we use the same parameters, $am$ and $\cal A$, as in the zero-field case above.
The agreement is very good 
since the corrections, $\Delta am$ and $\Delta{\cal A}$, associated with the external field 
($a^2 q_d {\cal E} = 10^{-2}$ in this case) are minute.
Interestingly, the wavefunction develops nodes between the walls in the presence of background electric field. It is tempting to think that this is due to the acceleration in the presence of the field, but in fact this
is due to the gauge choice for the external field.


\subsection{Polarizability extraction}
To determine the pion polarizbility, we computed the lattice QCD correlators $G(x,t)$ on the four ensembles in 
Table~\ref{tab:ens} for charged pions, and carried out correlated fits as described above. 
We have used the same fit region that we had used when determining $\tilde L$ of the effective model in 
Table~\ref{tab:ens}. Furthermore, for ensembles EN2, EN3, and EN4 which have the same temporal extension, 
we have fixed the temporal fit range to be the same for all of them. 

\begin{figure}[t]
    \centering
    \begin{tabular}{c}
    \parbox[c]{0.95\columnwidth}{\includegraphics[width=\linewidth,trim=0cm 0 0 0]{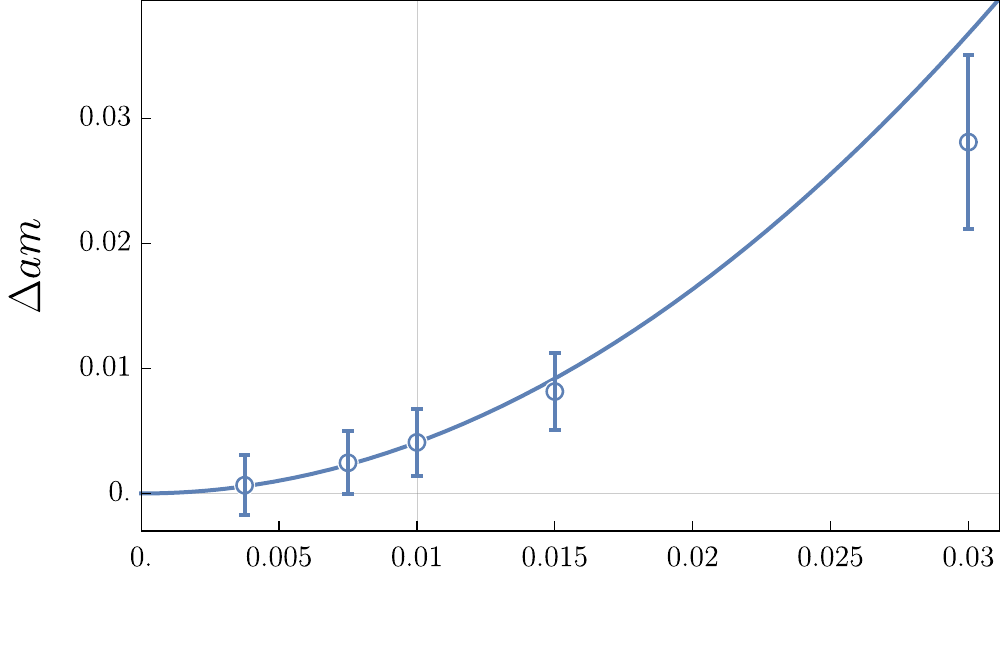} } \vspace{-0.cm} \\
    \parbox[c]{0.95\columnwidth}{\includegraphics[width=\linewidth,trim=0cm 0 0 0]{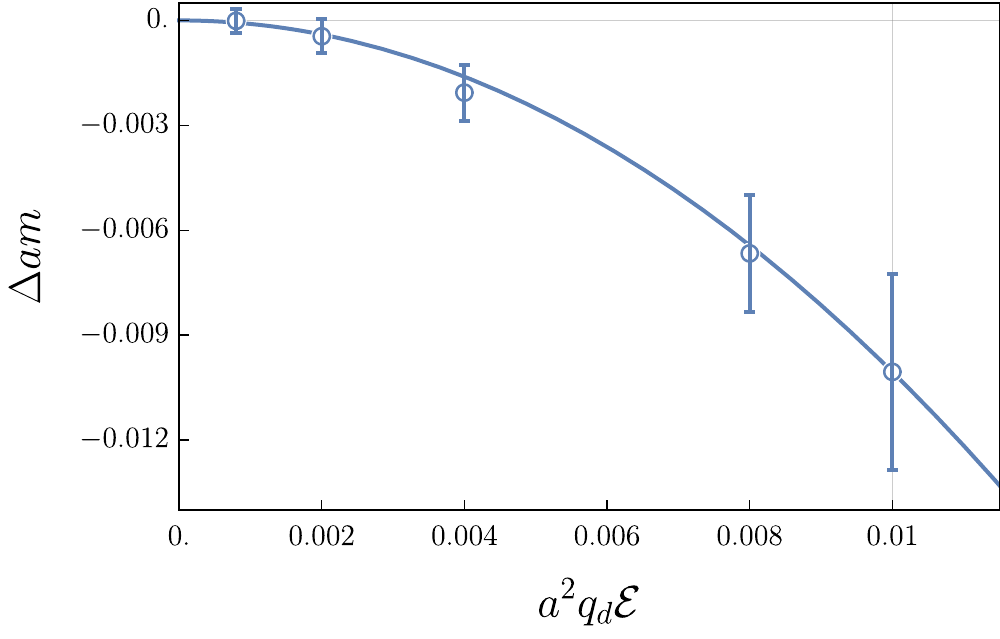} }
    \end{tabular}
    \caption{Mass shift as a function of the electric field for EN1 (top) and EN4 (bottom). The curves plotted here are the quadratic function which goes through zero and the central value of $\Delta am$ at $a^2q_d \mathcal{E} = 0.01$. The quadratic behavior breaks down for larger values of the electric field since the higher-order corrections to the energy of the pion dominate.
    }
    \label{fig:scaling}
\end{figure}
\begin{table}[b]
\setlength{\tabcolsep}{3.5pt}
\renewcommand{\arraystretch}{1.4}
	\centering
	\begin{tabular*}{0.85\columnwidth}{@{\extracolsep{\stretch{1}}}cccc@{}}
		\toprule
		Ensemble & $\Delta am$ & $\alpha_{\pi^\pm} \, [10^{-4} \fm^3]$ & $\chi^2/{\rm dof}$ \\ 
		\midrule
		EN1 & 0.0041(27) & $1.30(84)$ & 1.4  \\
		EN2 & 0.0013(14) & $0.41(43)$ & 1.6 \\
		EN3 & 0.0040(16) & $1.26(50)$ & 1.4 \\
		EN4 & -0.0101(28) & $-3.17(87)$ & 1.6 \\
		\bottomrule
	\end{tabular*}
	\caption{Mass shift in lattice units and extracted electric polarizability for charged pions for a pion mass of $315\MeV$.}
	\label{tab:res}
\end{table}

To make sure that we are extracting the quadratic response to the electric field, we compute the mass shift for a sequence of increasing values for the electric field strength.
In Fig.~\ref{fig:scaling} we plot the mass shift as a function of field strength for the smallest ensemble~(EN1) and the largest~(EN4). We see that the value of the electric field used in this study is indeed in the regime where the mass shift scales quadratically with the electric field strength. For our study we used the smallest value of the electric field where the mass-shift is not statistically compatible with zero on the smallest ensemble~(EN1): $a^2 q_d {\cal E}=0.01$.

The fit results are summarized in Table~\ref{tab:res}. The mass shifts are indeed small when compared to the particle mass. The conversion from lattice units to physical units from Eq.~\ref{eq:shift} is given by
\beq
\alpha_{\pi^\pm} = \frac{2}{a {\cal E}^2} \delta am
\eeq
where $a=0.1245$ fm, $e^2=1/137$, and $a^2 q_d{\cal E}=0.01$.

\begin{figure}[h!]
  \centering
  \includegraphics[width=0.9\columnwidth,trim= 0 0 0 +0.5cm]{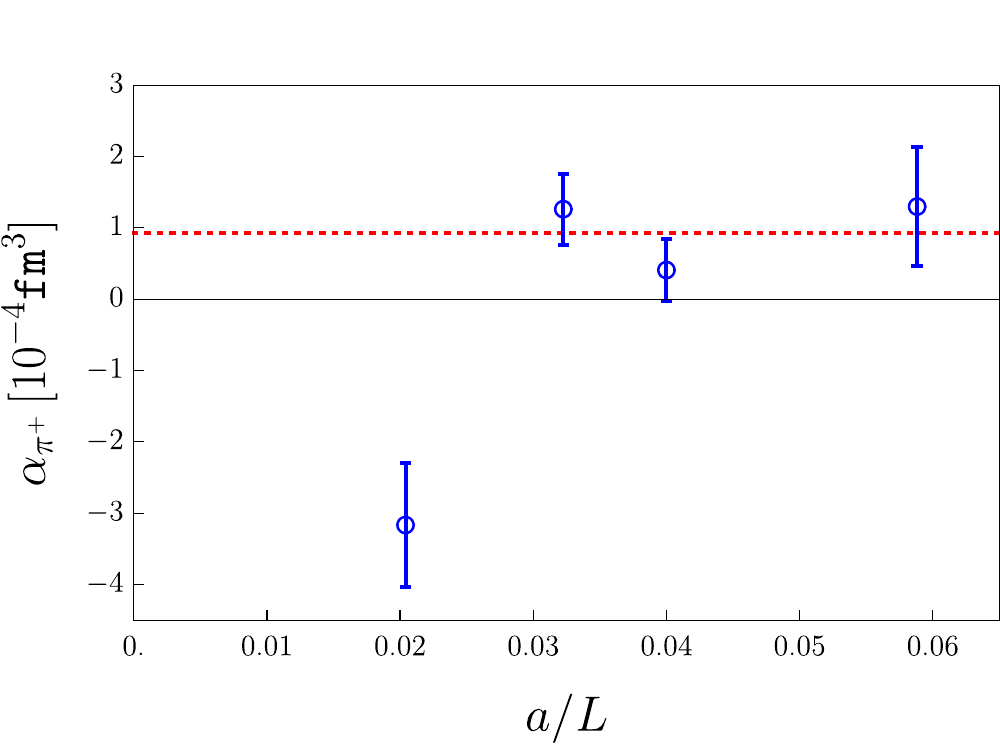}
  \caption{Volume dependence of the extract polarizability at pion mass of $m_\pi=315\MeV$. The dotted line is prediction from leading-order ChPT.}
  \label{fig:infvol}
\end{figure}

\comment{
To help resolve the volume dependence, we use a quantitative criterion called the Akaike Information Criterion (AIC)~\cite{1100705} which can tell us what is the best fit for our model that avoids overfitting. This quantity is given by
\beq
  \text{AIC}  =  2k + \chi^2 \, ,
\eeq
where $k$ is the number of parameters in our fit and $\chi^2$ is the total $\chi^2$ for that specific fit. The fit with the smallest AIC will be the best candidate. For a constant fit we have AIC=3.9 and for a linear one we have AIC=5.8. So, a constant volume dependence is favored; it describes the data well enough without overfitting. This can also be seen in Fig.~\ref{fig:infvol}. 
}
\comment{
Our final result for the electric polarizability of charged pions is 
\beq
  \alpha_{\pi^\pm} = 0.81(32)(xx) \times 10^{-4} \, {\fm}^3
  \label{eq:alpha}
\eeq
%
where the first uncertainty is statistical and the second is the aggregate of systematic uncertainty.
}

The dominant sources of systematic errors are the choice of the fit range in the spatial and temporal direction,
and the value of $\tilde L$ using in the effective model. To determine their contribution we analyze the
sensitivity of the polarizability to expected changes in these parameters. For the fitting range, we reduced
both the $x$ and $t$-ranges by one unit of lattice spacing. For $\tilde L$ we vary it within the error bands
reported in Table~\ref{tab:ens}.
The results are included in Table~\ref{tab:sensitivity}. 
We note here that for the three largest ensembles used in our study the finite-volume corrections estimate from
ChPT is less than 10\%~\cite{Tiburzi:2008pa}.

%
%
%
\begin{table}[b]
\setlength{\tabcolsep}{3.5pt}
\renewcommand{\arraystretch}{1.4}
	\centering
	\begin{tabular*}{0.95\columnwidth}{@{\extracolsep{\stretch{1}}}ccccccc@{}}
		\toprule
		Ensemble &\phantom{}& Original &&  \renewcommand\arraystretch{1.1}\begin{tabular}{@{}c@{}}$x_i+1$ \\ $x_f-1$\end{tabular} & $t_i+1$ & $\tilde L-\delta \tilde{L}$ \\ 
		\midrule
		EN1 && 1.30(84) && 1.22(85)  & 1.18(87) & 1.57(85) \\
		EN2 && 0.41(43) && 0.35(43)  & 0.51(47) & 0.68(43) \\
		EN3 && 1.26(50) && 1.26(50)  & 1.34(54) & 1.50(50) \\
		EN4 && -3.17(87)&& -3.13(85) & -4.4(1.2) & -3.00(87)\\
		\bottomrule
	\end{tabular*}
	\caption{Sensitivity of the results on the choice of fit ranges and the model size. We compare the polarizability $\alpha_{\pi^\pm}$ extracted using our choice for the fit ranges with the value extracted for slightly different choices. In this table $\delta{\tilde L}$ is the error on $\tilde L$ reported in Table~\ref{tab:ens}.}
	\label{tab:sensitivity}
\end{table}

In Fig.~\ref{fig:infvol} we plot the
results for polarizability as a function of
the lattice extent in the field direction.
Our results are compared with the ChPT value of~\cite{Holstein:2013kia}
\beq
  \alpha_{\pi^\pm} = \frac{\alpha_{\text{em}}}{8\pi^2 m_\pi f_\pi^2} \frac{h_A}{h_V} = 0.93(6) \times 10^{-4} \, {\fm}^3,
  \label{eq:ChPT}
\eeq
evaluated at our pion mass of $m_\pi=315\MeV$. To compute the pion decay constant for our pion mass we used NLO ChPT
expression with low-energy constant $\overline{\ell}_4$ from the FLAG review~\cite{Aoki:2019cca}  
to get $f_\pi=106\MeV$. For the form factors ratio $h_A/h_V$ we used the experimental value $ 0.469\pm0.031$~\cite{Holstein:2013kia}
which also gives the dominant source of error to the ChPT estimate.
For the smallest ensembles, our results are in agreement with the ChPT estimate. For the largest volume~(EN4) the polarizability turns, surprisingly, negative. We do not have an explanation for this behavior but we note that there are other lattice calculations of pion polarizability for comparable quark masses that are at variance with ChPT expectations. One is on magnetic polarizability that is expected to satisfy $\beta_{\pi^+} = -\alpha_{\pi^+}$, a calculation for $m_\pi=296\MeV$ finds $\beta_{\pi^+}=0.35(11)\times 10^{-4}\fm^3$. 
Another calculation of the ``connected" neutral pion electric polarizability---where the disconnected diagrams are not included---finds that the polarizability turns negative for $m_\pi<350\MeV$~\cite{Lujan:2014kia,Lujan:2016ffj} although the ChPT expectations are that this value should be positive~\cite{Detmold:2009dx}.

\section{\label{sec:con} Conclusion}

Computing electric polarizability for charged hadrons has been a challenge for lattice QCD due to the acceleration 
experienced by charged particles in the background electric field. Using the lightest charged hadron, the pion, as an example, 
we presented a method to extract the electric polarizability. 
The method relies on an effective field model that is designed to capture the behavior of the charged pion in a box with the 
same geometry as that used in lattice simulations. This has proven crucial since the infinite-volume version of the 
model fails to account for the significant finite-volume effects. This is important to isolate the energy shift due only 
to the deformation of the hadron from the effects due to its motion in the electric field.

To fit the lattice QCD correlators we need to compute its $x$ and $t$ dependence separately to match the effective model. 
We construct a $\chi^2$-function that utilizes information in both the real and imaginary parts of the correlator simultaneously 
and is invariant under gauge transformations of the background field.

Since we use Dirichlet boundary conditions, we are not limited to just quantized values under periodic boundary conditions. 
We can dial the electric field continuously and choose values that give better signals.

We extract the pion polarizability on a set of ensembles with different distance between the Dirichlet walls. For the pion mass $m_\pi=315\MeV$ used in this study, the results on the smallest three ensembles are compatible with a small positive value for $\alpha$, a value in agreement with ChPT predictions. However, on the largest ensemble the polarizability is negative, a puzzling result. We have checked the sensitivity of the polarizability on the fitting range and the effective distance between the walls employed in the effective model. The effect of varying these parameters is smaller than the stochastic error and cannot explain the shift to negative values for the polarizability. We note that there are other lattice calculations of electromagnetic polarizabilities at similar pion mass that in tension with ChPT predictions.

In this paper we focused more on the technical aspects of the extraction method. Looking to the future, 
in order to compare with experiments, we need to extrapolate to physical quark masses. We expect that as we approach the physical mass ChPT predictions become more reliable and should offer a strong check for our results. In particular we plan to investigate the puzzling result we got on the largest volume in this study. 
ChPT predicts a $1/m_\pi$ rise as the pion mass is lowered (see Eq.~\ref{eq:ChPT}) so the signal might be stronger at smaller pion masses. 
Another direction we plan to investigate is the effect of lattice discretization. As an additional check of the fitting procedure, 
we plan to apply the same method to neutral mesons on the same set of ensembles where simple exponential fits were
previously employed~\cite{Lujan:2016ffj, Lujan:2014kia}. Another systematic effect, namely charging the sea quarks, 
is not addressed here, though such effects are expected to be small~\cite{Freeman:2014kka, Freeman:2013eta}.
In the long run, an extension to baryons would be desirable with an eye towards the proton electric polarizability 
which is more precisely measured than charged pions but is less well studied on the lattice.

\bigskip
\begin{acknowledgments}
This work was supported in part by DOE Grant~No.~DE-FG02-95ER40907. AA gratefully acknowledges the hospitality of the 
Physics Department at the University of Maryland where part of this work was carried out. The computations were performed 
on the GWU Colonial One computer cluster and the GWU IMPACT collaboration clusters.
\end{acknowledgments}

\bibliographystyle{JHEP}
\bibliography{main}

\end{document}